\documentclass{emulateapj}
\usepackage{color}

\slugcomment{draft v2}

\shorttitle{Scattered sGRBs}
\shortauthors{Kisaka et al.}

\begin{document}


\title{Scattered Short Gamma-Ray Bursts as Electromagnetic Counterparts to Gravitational Waves and Implications of GW170817 and GRB 170817A}


\author{Shota Kisaka\altaffilmark{1,6}}
\email{kisaka@phys.aoyama.ac.jp}
\author{Kunihito Ioka\altaffilmark{2}}
\email{kunihito.ioka@yukawa.kyoto-u.ac.jp}
\author{Kazumi Kashiyama\altaffilmark{3,4}}
\email{kashiyama@phys.s.u-tokyo.ac.jp}
\author{Takashi Nakamura\altaffilmark{2,5}}
\email{nakamura.takashi.57a@st.kyoto-u.ac.jp}


\altaffiltext{1}{Department of Physics and Mathematics, Aoyama Gakuin University, Sagamihara, Kanagawa, 252-5258, Japan}
\altaffiltext{2}{Center for Gravitational Physics, Yukawa Institute for Theoretical Physics, Kyoto University, Kyoto 606-8502, Japan}
\altaffiltext{3}{Department of Physics, the University of Tokyo, Bunkyo, Tokyo 113-0033, Japan}
\altaffiltext{4}{Research Center for the Early Universe, the University of Tokyo, Tokyo 113-0033, Japan}
\altaffiltext{5}{Department of Physics, Kyoto University, Kyoto 606-8502, Japan}
\altaffiltext{6}{JSPS Research Fellow}


\begin{abstract}
In the faint short gamma-ray burst sGRB 170817A 
followed by the gravitational waves (GWs) from a merger of two neutron stars (NSs) GW170817, 
the spectral peak energy is too high to explain only by canonical off-axis emission. 
We investigate off-axis appearance of an sGRB prompt emission scattered 
by a cocoon, which is produced through the jet-merger-ejecta interaction,
with either sub-relativistic or mildly-relativistic velocities.
We show that the observed properties of sGRB 170817A, 
in particular the high peak energy, 
can be consistently explained by the Thomson-scattered emission with a typical sGRB jet,
together by its canonical off-axis emission,
supporting that an NS-NS merger is the origin of sGRBs.
The scattering occurs at $\lesssim 10^{10}$--$10^{12}\,{\rm cm}$ not far from the central engine,
implying the photospheric or internal shock origin of the sGRB prompt emission.
The boundary between the jet and cocoon is sharp, which could be probed by future observations
of off-axis afterglows. 
The scattering model predicts 
a distribution of the spectral peak energy that is similar to the observed one 
but with a cutoff around $\sim$ MeV
energy, and its correlations with 
the luminosity, duration, and time lag from GWs,
providing a way to distinguish it from alternative models. 
\end{abstract}





\section{INTRODUCTION}

On 17 August 2017, gravitational waves (GWs) from a merger of two neutron stars (NSs)
were finally discovered by
the Laser Interferometer Gravitational-Wave Observatory (LIGO) along with Virgo,
and dubbed as GW170817 \citep{2017PhRvL.119p1101A}.
{\it Fermi} Gamma-ray Burst Monitor ({\it Fermi}/GBM) and {\it INTEGRAL} 
also detected a short (duration $\sim 2$ s) gamma-ray burst sGRB 170817A 
$\sim 1.7$ s after the GWs ceased
\citep{2017ApJ...848L..13A, 2017arXiv171005449S, 2017arXiv171005446G}.
Follow-up observations involving more than 3000 people identified the counterparts
across the electromagnetic (EM) wavelengths \citep{2017ApJ...848L..12A},
which also spotted the E/S0 host galaxy NGC 4993 about $40$ Mpc away \citep{2017ApJ...848L..31H, 2017ApJ...849L..16I}.
This event is really the historical breakthrough of the multi-messenger astronomy.

An NS binary merger is long thought to be
the most likely origin of sGRBs \citep[e.g., ][]{1986ApJ...308L..43P, 1989Natur.340..126E, 1992ApJ...397..570M, 1992ApJ...395L..83N}.
The observational results such as a wide variety of their host galaxy type and an absence of associate supernovae 
are consistent with the merger scenario, but not conclusive evidence \citep[e.g., ][]{2014ARA&A..52...43B}. 
Simultaneous detections of GWs and an sGRB should be
a smoking-gun evidence for the merger scenario.
This time, the apparent isotropic-equivalent energy
$E_{\rm iso} \sim 5 \times 10^{46}$ erg is significantly 
lower than that of ordinary sGRBs.
This is also expected to some extent because 
an sGRB is caused by a relativistic jet 
with the emission beamed into a narrow solid angle.
The GW observations allow the viewing angle 
$\theta_{\rm v} \lesssim 32^{\circ}$
relative to the total angular momentum axis
\citep{2017PhRvL.119p1101A, 2017Natur.551...85A}.
Thus sGRB 170817A is likely the first off-axis sGRB ever observed.

Actually an off-axis jet of a canonical sGRB
seems to be consistent with all the EM signals currently observed in GW170817 
\citep{2017ApJ...848L..13A, 2017arXiv171005905I, 2017ApJ...850L..24G}.
The off-axis de-beaming of the emission can easily decrease the apparent isotropic energy 
down to the observed value $\sim 5 \times 10^{46}$ erg,
where we should be careful that the de-beaming is less significant than the point source case because
the opening angle of the jet is comparable to the viewing angle
\citep{2017arXiv171005905I}.
During the propagation in the material ejected at the merger,
the jet also injects energy into the cocoon, 
and the jet-powered cocoon could be observed as 
the detected blue macronova (or kilonova) at $\sim 1$ day
\citep{2017ApJ...848L..12A, 2017PASJ...69..102T, 2017PASJ...69..101U, 2018PASJ..tmp...22T, 2017Natur.551...75S, 
2017Natur.551...64A, 2017Natur.551...80K, 2017Sci...358.1570D, 2017Sci...358.1574S, 2017Sci...358.1559K, 2017Sci...358.1583K, 2017Sci...358.1556C, 2017Natur.551...67P, 2017ApJ...848L..17C, 2017ApJ...848L..18N, 2017ApJ...848L..26S, 2017ApJ...848L..16S, 2017ApJ...850L...1L, 2017SciBu..62.1433H, 2017ApJ...848L..29D, 2017PASA...34...69A, 2017NatAs...1..791C, 2017ApJ...848L..32M, 2017ApJ...848L..24V, 2018MNRAS.474L..71B, 2018arXiv180207732M}.
The jet is finally decelerated by the interstellar medium,
leading to the off-axis afterglow emission 
that can fit both the observed X-ray and radio signals
\citep{2017Natur.551...71T, 2017Sci...358.1565E, 2017Sci...358.1579H, 2017arXiv171008514F, 2017ApJ...848L..20M, 2017ApJ...848L..21A, 2017ApJ...848L..25H, 2017ApJ...850L..21K}.
Such an off-axis jet model is the most simple and may be preferred by Occam's razor.

Nevertheless, the low luminosity of the prompt emission might be
explained by other scenarios, such as 
emission from a structured jet with a wide-angle distribution \citep{2016ApJ...829..112L, 2017arXiv170807008J, 2017arXiv171006421G, 2017ApJ...850L..41X, 2018NatCo...9..447Z, 2017arXiv171005823B},
breakout emission from a mildly-relativistic cocoon
\citep{2017arXiv171005896G, 2018ApJ...852L..30P, 2018ApJ...855..103P}
and on-axis emission from a low-luminosity sGRB population
\citep{2017ApJ...848L..34M, 2017arXiv171005869H, 2018ApJ...853L..10Y, 2018ApJ...852L...1Z, 2017arXiv171005857L}.
Some of these scenarios will be tested by the future radio and X-ray observations.

One possible tension against the off-axis jet scenario
is the peak energy of the sGRB spectrum.
Most of the currently proposed models also suggest
a relatively soft peak energy, $\sim 1-10$ keV \citep[e.g., ][]{2017arXiv171005905I, 2018ApJ...852L..30P, 2017ApJ...851L..19B}.
From the detail analysis of the {\it Fermi}/GBM data, 
the observed peak energy of the main pulse is 
$E_{\rm p} \sim 185 \pm 62$ keV 
for a Comptonized spectrum (a power law with an exponential cutoff)
\citep{2017arXiv171005446G, 2017arXiv171008362B},
while the weak tail has a blackbody spectrum with $kT = 10.3 \pm 1.5$ keV.
Of course, we should be careful about these values
because the observed flux $\sim 1.9$ photons cm$^{-2}$ s$^{-1}$ is just above
the detection threshold of {\it Fermi}/GBM $\sim 0.5$ photons cm$^{-2}$ s$^{-1}$
\citep{2017GCN.21520....1V}.
Indeed the peak energy is shifted softward to 
$E_{\rm p} \sim 70$ keV
if the spectrum is fit by the Band function \citep{2017ApJ...848L..13A}. 
In addition the peak energy $E_{\rm p} \sim 185 \pm 62$ keV 
is still consistent with any low peak energies within $3 \sigma$.
Some analyses seem to adopt a prior that does not allow a low peak energy ($\lesssim10$ keV). 
Nevertheless, the obtained peak energy could imply a different emission mechanism, 
which may be also working together with the off-axis de-beamed emission.
This is worth to explore for future observations.

\begin{figure*}
 \begin{center}
  \includegraphics[width=140mm]{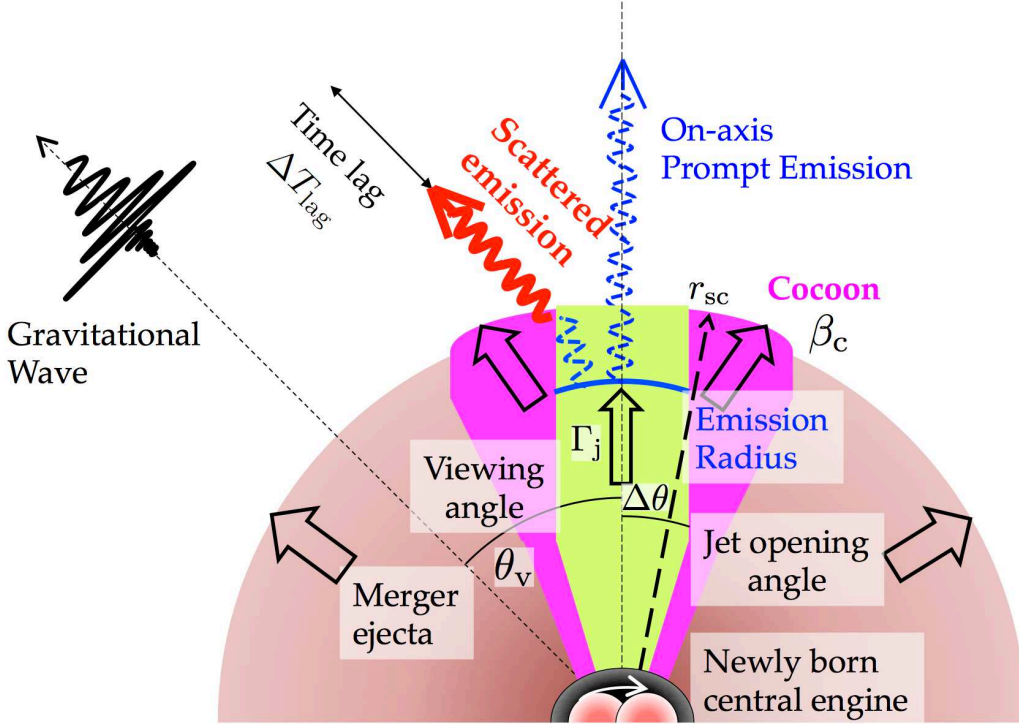}
   \caption{Schematic picture for the scattered prompt sGRB.}
  \label{fig:image}
 \end{center}
\end{figure*}

In this paper, we propose the Thomson scattering of the prompt emission as an EM counterpart to binary NS mergers. 
Scattering in GRBs has been discussed as a mechanism to make
wide-angle emission \citep[e.g., ][]{1998PThPh.100..921N, 1999ApJ...521L.117E, 2015ApJ...809L...8K}.
Figure \ref{fig:image} shows the schematic picture for 
scattering of prompt emission in a binary NS merger. 
At the merger of the NSs, strong GWs are emitted ($t=0$). 
After the merger, a part of the NS mass is ejected \citep[e.g., ][]{2013PhRvD..87b4001H}, and
a relativistic jet is launched from the central compact object.  
During the propagation in the merger ejecta, 
a part of the jet energy is injected into the cocoon.
After the jet penetrates the merger ejecta, the photons in the jet can escape. 
If the viewing angle of $\theta_{\rm v}$ relative to the jet axis is smaller than the jet opening angle $\Delta\theta$, 
the escaping photons would be observed as the prompt emission of an sGRB. 
On the other hand, photons scattered by the cocoon could dominate the $\gamma$-ray flux for off-axis observers.

The scattering model can naturally explain a peak energy in the main pulse 
as high as those of canonical sGRBs since the scattered photon energy is 
similar to the unscattered one
unless the photon energy is $\gtrsim1$ MeV. 
If the jet opening angle $\Delta\theta$ is wider than the beaming angle $1/\Gamma_{\rm j}$, 
where $\Gamma_{\rm j}$ is the bulk Lorentz factor of the jet, 
a fraction $2/\Gamma_{\rm j} \Delta\theta$ 
(e.g., the ratio of a ring $\sim 2\pi \Delta\theta/\Gamma_{\rm j}$
to a circle $\sim \pi \Delta\theta^2$ of the solid angle) 
of the on-axis emission
can be scattered at the jet-cocoon boundary to $\sim 4\pi$ directions.
Then the isotropic luminosity of the scattered emission is
\begin{eqnarray}\label{eq:order}
L_{\rm sc}&\approx&
\epsilon_{\rm sc}
\frac{2}{\Gamma_{\rm j}\Delta\theta}\frac{\Delta\theta^2}{2}L_{\rm iso} \nonumber \\
&\approx&3\times10^{46}\,{\rm erg~s}^{-1}
\left(\frac{\epsilon_{\rm sc}}{0.1}\right)
\left(\frac{\Delta\theta}{0.3}\right)
\nonumber\\&\times&
\left(\frac{\Gamma_{\rm j}}{10^3}\right)^{-1}
\left(\frac{L_{\rm iso}}{10^{51}~{\rm erg~s}^{-1}}\right),
\end{eqnarray}
where $(\Delta\theta^2/2)L_{\rm iso}$ is the geometrically-corrected luminosity 
of the on-axis prompt emission,
and $\epsilon_{\rm sc}$ is a correction factor depending on details
(see next section).
Interestingly the above rough estimate is comparable to 
the observed luminosity of sGRB 170817A.
Note that both the emissions of scattering and off-axis de-beaming can occur in a single sGRB,
and they complement each other.

In this paper, we consider the Thomson scattering of the sGRB prompt emission.
In Sec.~\ref{sec:scattered_prompt},
we generally formulate the scattering mechanism 
to calculate the luminosity, duration, and time lag
of the scattered emission for an off-axis observer with
$\theta_{\rm v}>\Delta\theta$.
Then, in Sec.~\ref{sec:scatterer}, 
we apply the formulae to an sGRB prompt emission from a binary NS merger.
We consider a non-relativistic scatterer or cocoon in Sec.~\ref{sec:nonrela},
and a relativistic scatterer, 
which may be 
the head part of the merger ejecta produced at the onset of the merger \citep{2014MNRAS.437L...6K}
or a mildly-relativistic cocoon 
due to the imperfect mixing of the jet and the merger ejecta \citep[e.g., ][]{2017ApJ...834...28N, 2018MNRAS.473..576G, 2017ApJ...848L...6L},
in Sec.~\ref{sec:rela}.
In Sec.~\ref{sec:spec},
we again emphasize that the scattering model generally predicts
the spectral peak energy $E_{\rm p}$ similar to the on-axis value
if $E_{\rm p} \lesssim {\rm MeV}$.
Finally, we apply the scattering model to GRB 170817A/GW170817 and discuss the implications in Sec.~\ref{sec:implications}. 
In Sec.~\ref{sec:jet-cocoon}, we discuss the constraints from the observations such as afterglow emission.
We present the summary and discussions in Sec.~\ref{sec:discussion}. 
Hereafter, we use $Q_x\equiv Q/10^x$ in cgs units. 

\section{General formulation of scattering}
\label{sec:scattered_prompt}

We consider prompt emission at a radius $r_{\rm emi}$ 
from a relativistic jet 
with a Lorentz factor $\Gamma_{\rm j}=1/\sqrt{1-\beta_{\rm j}^2}$ 
and an opening angle $\Delta\theta \gg \Gamma_{\rm j}^{-1}$.
The observed isotropic luminosity for an on-axis observer is $L_{\rm iso}=4\pi d_{\rm L}^2F_{\rm obs}$, 
where $d_{\rm L}$ is the luminosity distance and $F_{\rm obs}$ is the observed flux. 
The observed duration of the prompt emission for an on-axis observer is usually 
determined by the intrinsic engine activity timescale $t_{\rm dur}$. 
The jet is surrounded by a scatterer with an expansion velocity of $\beta_{\rm c}c$, 
either relativistic or non-relativistic.
In order to scatter the prompt emission, 
the optical depth of the scatterer has to be order of unity
\begin{eqnarray}\label{eq:tau}
\tau \sim 1,
\end{eqnarray}
and the emission radius has to be comparable or
smaller than the scattering radius, 
\begin{equation}\label{eq:r_emi}
r_{\rm emi} \lesssim r_{\rm sc}.
\end{equation}

The isotropic luminosity of the scattered emission $L_{\rm sc}$ is 
obtained from the geometrically-corrected luminosity 
$(\Delta\theta^2/2)L_{\rm iso}$ as
\begin{equation}\label{eq:L_sc,iso}
L_{\rm sc} \approx \frac{2}{\Gamma_{\rm j}\Delta\theta}
\times\frac{t_{\rm dur}}{T_{\rm dur, sc}}\times\Gamma_{\rm c}^2
\times \epsilon_{\rm sc} \times \frac{\Delta\theta^2}{2} L_{\rm iso}.
\end{equation}
Here (a) the first factor $2/\Gamma_{\rm j}\Delta\theta$
is the fraction of the geometrically-corrected luminosity $(\Delta\theta^2/2)L_{\rm iso}$
that can collide with the scatterer at the scattering radius,
and is given by the ratio of the solid angles between the jet $\pi \Delta\theta^2$
and an outer ring $\sim 2\pi \Delta\theta/\Gamma_{\rm j}$
with a width of the relativistic beaming angle $\sim 1/\Gamma_{\rm j}$.
(b) The second factor is the ratio of the on-axis duration $t_{\rm dur}$ and
the off-axis duration of the scattered emission $T_{\rm dur,sc}$.
The observed duration of the scattered emission is given by
\begin{equation}\label{eq:T_sc,emi}
T_{\rm dur, sc} \approx \max\left[t_{\rm dur}, \Delta T \right], 
\end{equation}
which may be longer than the jet activity timescale $t_{\rm dur}$
because the duration of a single scattered pulse $\Delta T$ 
is determined by the smaller Lorentz factor of the scatterer $\Gamma_{\rm c}$ than $\Gamma_{\rm j}$ as
\begin{eqnarray}\label{eq:T_sc}
\Delta T &\sim& \frac{r_{\rm sc}}{c\beta_c}[1-\beta_{\rm c}\cos(\theta_{\rm v}-\Delta\theta)] \nonumber \\
&\approx& \frac{r_{\rm sc}}{2c\beta_{\rm c}\Gamma_{\rm c}^2},
\end{eqnarray}
in the case of $\Gamma_{\rm c}^{-1}>\theta_{\rm v}-\Delta\theta$.
Note that the scattering occurs over a range of radii around $r_{\rm sc}$,
not a thin shell at $r_{\rm sc}$.
The total energy of $(2/\Gamma_{\rm j}\Delta\theta)(\Delta\theta^2/2)L_{\rm iso} t_{\rm dur}$ 
is emitted with a duration of $T_{\rm dur, sc}$,
so that the luminosity is reduced by the factor of $t_{\rm dur}/T_{\rm dur,sc}$.
(c) The third factor $\Gamma_{\rm c}^2$ is the relativistic beaming factor 
for a relativistic scatterer because 
the scattered emission is beamed into a cone with an opening angle $\sim 1/\Gamma_{\rm c}$.  
(d) The last factor $\epsilon_{\rm sc}$ is a correction factor coming from, e.g.,
the ratio between the emission and scattering radii, $r_{\rm emi}/r_{\rm sc}$,
the opacity of the scatterer, and so on.
For example, \citet{1999ApJ...521L.117E} give $\epsilon_{\rm sc}\sim10^{-3}-10^{-1}$ for
$r_{\rm emi}/r_{\rm sc}\sim0.1$. 
Most of the energy is scattered within the angle, 
\begin{equation}\label{eq:opening}
\Delta \theta_{\rm sc} \approx \frac{1}{\Gamma_{\rm c}}.
\end{equation}

The jet may be collimated by the cocoon pressure. 
In the collimated case, the Lorentz factor of the jet during the propagation in the merger ejecta is suppressed 
\citep[e.g., ][]{2011ApJ...740..100B, 2013ApJ...777..162M}. 
Since the scattered fraction is proportional to the beaming angle $\sim1/\Gamma_{\rm j}$ as in Eq. (\ref{eq:L_sc,iso}), 
the scattered luminosity $L_{\rm sc}$ could be enhanced. 

The time lag of the scattered emission behind the GW signal is given by
\begin{equation}\label{eq:t_lag}
\Delta T_{\rm lag} \approx t_{\rm j}+t_{\rm br}+\Delta T_{\rm scbr},
\end{equation} 
where $t_{\rm j}$ is the time of the jet launch after the merger,
and $t_{\rm br}$ is the timescale for the jet breakout from the merger ejecta\footnote{
More precisely, the observed breakout time is given by
$t_{\rm br}[1-\beta_{\rm h}\cos(\theta_v-\Delta\theta)]$,
where $\beta_{\rm h}$ is the jet head velocity during the propagation in the merger ejecta.
Since $\beta_{\rm h}$ is usually non-relativistic \citep{2017arXiv171005905I},
the approximation of using $t_{\rm br}$ is adequate.
}.
The last term in Eq. (\ref{eq:t_lag}) is the delay time of a single pulse 
caused by the curvature effect of the emission surface 
at the scatter-starting radius $r_{\rm scbr}$, 
\begin{eqnarray}\label{eq:T_scbr}
\Delta T_{\rm scbr}&\sim& \frac{r_{\rm scbr}}{c\beta_c}[1-\beta_{\rm c}\cos(\theta_{\rm v}-\Delta\theta)] \nonumber \\
&\approx& \frac{r_{\rm scbr}}{2c\beta_{\rm c}\Gamma_{\rm c}^2},
\end{eqnarray}
in the case of $\Gamma_{\rm c}^{-1}>\theta_{\rm v}-\Delta\theta$.
The jet can be launched either within a dynamical timescale after the merger
($t_{\rm j}\sim1-10$ ms) or after a hyper-massive NS, if any, collapses into a BH 
($t_{\rm j}\gtrsim 0.1$ s).
The jet breakout timescale should be $t_{\rm br}\lesssim t_{\rm dur}$ for a successful breakout \citep{2014ApJ...784L..28N}.

In summary, for a given set of parameters of the prompt emission 
($r_{\rm emi}$, $r_{\rm sc}$, $r_{\rm scbr}$, $L_{\rm iso}$, $\Gamma_{\rm j}$, 
$\Delta\theta$, $t_{\rm dur}$, $t_{\rm j}$, $t_{\rm br}$, $\beta_{\rm c}$) 
and a viewing angle $\theta_{\rm v}$, 
the properties of the scattered emission ($L_{\rm sc}$, $T_{\rm dur, sc}$, $\Delta T_{\rm lag}$) 
can be calculated with Eqs.~(\ref{eq:r_emi})--(\ref{eq:t_lag}).
In the next section,
we setup the parameters in the context of an sGRB from a binary NS merger.

\section{Applications to scattered sGRBs}
\label{sec:scatterer}

\subsection{Non-relativistic scatterer}\label{sec:nonrela}

First, we consider scattering by non-relativistic matter, namely, non-relativistic cocoon.
During the propagation of the jet in the merger ejecta, a part of the energy is 
injected to the cocoon surrounding the jet, which can scatter the prompt emission. 
If the shocked jet and merger ejecta are efficiently mixed before the jet-cocoon breakout, 
the expansion velocity of the cocoon becomes sub-relativistic, 
$\sim0.2$--$0.4c$ \citep{2017arXiv171005905I}, 
which is automatically adjusted to be slightly faster than the
typical velocity of the dynamical ejecta \citep[$\sim0.1$--$0.2c$; ][]{2013PhRvD..87b4001H, 2015PhRvD..91f4059S, 2016CQGra..33r4002L, 2016MNRAS.460.3255R, 
2016PhRvD..93d4019F, 2017PhRvD..95d4045D, 2017PhRvD..95b4029D, 2017PhRvD..96l4005B, 2017PhRvD..96l3012S}. 

The scattering radius is around the outer radius of the cocoon
where the optical depth for the Thomson scattering is order of unity in Eq.~(\ref{eq:tau}).
Since the irradiated cocoon is fully ionized near the jet-cocoon boundary,
the electron number density of the cocoon is roughly
$n\sim 3M_{\rm c}/(8\pi m_{\rm p}\beta_{\rm c}^3c^3t^3f_{\rm cv})$, 
where $M_{\rm c}=f_{\rm cm}M_{\rm e}$ is the cocoon mass, 
$M_{\rm e}=10^{-2}M_{\rm e,-2}M_{\odot}$ is the ejecta mass, 
$f_{\rm cm}$ is the mass fraction of the cocoon, 
and $f_{\rm cv}$ is the fractional volume of the cocoon.
Typical values of the fraction are $f_{\rm cm}\sim f_{\rm cv}\sim0.5$ \citep{2017arXiv171005905I}. 
The optical depth for the Thomson scattering is high enough,
\begin{eqnarray}\label{eq:opticaldepth}
\tau\sim10^{10}t_{0}^{-2}(f_{\rm cm}/f_{\rm cv})M_{\rm e,-2}\beta_{\rm c,-0.5}^{-2},
\end{eqnarray}
which quickly drops below unity outside the cocoon radius.
Since the scattering can start at the breakout time $t=t_{\rm j}+t_{\rm br}$
and end at $t=t_{\rm j}+t_{\rm br}+t_{\rm dur}$,
the scattering radius ranges from
\begin{eqnarray}\label{eq:radius0}
r_{\rm scbr} &\sim& \beta_{\rm c}c (t_{\rm j}+t_{\rm br}),
\end{eqnarray}
to
\begin{eqnarray}\label{eq:radius}
r_{\rm sc} &\sim& \beta_{\rm c}c (t_{\rm j}+t_{\rm br}+t_{\rm dur})
\nonumber\\
&\sim& 1\times10^{10}\,{\rm cm}\,\beta_{\rm c,-0.5} (t_{\rm j}+t_{\rm br}+t_{\rm dur})_{0}.
\end{eqnarray}

In order to scatter the prompt emission, 
the emission radius has to be smaller than
the above scattering radii $r_{\rm sc}$ as in Eq.~(\ref{eq:r_emi}),
and this can be realized by the photospheric prompt emission.
The photospheric radius is determined by the optical depth, $n_{\rm j}'\sigma_{\rm T}r_{\rm ph}'\sim1$, 
where $n_{\rm j}'$ and $r_{\rm ph}'$ are the baryon number density and the photospheric radius 
in the jet comoving frame, respectively.
The mass conservation equation gives the number density, 
$n_{\rm j}'\sim L_{\rm iso}/(4\pi m_{\rm p}c^3r_{\rm ph}^2\eta\Gamma_{\rm j})$,
where $\eta$ is the photon-to-baryon ratio, $\sigma_{\rm T}$ is the Thomson cross section, 
and $m_{\rm p}$ is the proton mass.
Then, the photospheric emission radius of the jet 
can be estimated as \citep{2000ApJ...530..292M, 2006RPPh...69.2259M} 
\begin{eqnarray}\label{eq:photospheric}
r_{\rm emi} \sim r_{\rm ph} 
&\sim& \frac{L_{\rm iso}\sigma_{\rm T}}{4\pi m_{\rm p}c^3\eta\Gamma_{\rm j}^2} \nonumber \\
&\sim& 1\times10^{9}\,{\rm cm}\, \eta_3^{-1} L_{\rm iso, 51} \Gamma_{\rm j, 3}^{-2}.
\end{eqnarray}
Note that $\eta (\ge \Gamma_{\rm j})$ is fairly unknown and the key parameter to understand the GRB physics \citep[e.g., ][]{2000ApJ...530..292M, 2011PThPh.126..555I}. 
From Eqs. (\ref{eq:radius}) and (\ref{eq:photospheric}), 
two radii are comparable within reasonable ranges of parameters,
in particular for the case of a high Lorentz factor $\eta \ge \Gamma_{\rm j} \gtrsim 500$.
Therefore we consider $r_{\rm emi}\sim r_{\rm sc}$ in the non-relativistic case.
We note that the $e^{\pm}$ pair production opacity for 
$\gamma$-rays with the peak energy $E_{\rm p} \sim 100\,E_{\rm p,2}$ keV
is sufficiently small at the photosphere $r_{\rm emi}\sim10^{10}$ cm 
\citep[e.g., ][]{2001ApJ...555..540L, 2001ApJ...559..110Z, 2008ApJ...676.1123M}
\begin{eqnarray}\label{eq:compactness}
\tau_{\gamma\gamma}&\sim&\frac{L_{\rm iso}\sigma_{\rm T}}{4\pi E_{\rm p}cr_{\rm emi}\Gamma_{\rm j}^2}
\left(\frac{\Gamma_{\rm j} m_{\rm e} c^2}{E_{\rm p}}\right)^{2(1-\beta_{\rm B})}
\nonumber\\
&\sim& 10^{-5} L_{\rm iso,51}\Gamma_{\rm j,3}^{-5}r_{\rm emi,10}^{-1}E_{\rm p,2}^{2},
\end{eqnarray}
where $\beta_{\rm B}\sim2.5$ is the high-energy power-law index of the Band function \citep{1993ApJ...413..281B}, 
and $m_{\rm e}$ is the electron mass.

At the scattering radius in Eq.~(\ref{eq:radius}),
the duration of a single scattered pulse in Eq.~(\ref{eq:T_sc}) is
\begin{eqnarray}
\Delta T \sim 1\,{\rm s}\,(t_{\rm j}+t_{\rm br}+t_{\rm dur})_{0},
\end{eqnarray}
which is larger than the engine duration $t_{\rm dur}$
and hence the observed duration of the scattered emission $T_{\rm dur,sc}$
in Eq.~(\ref{eq:T_sc,emi}) is given by
\begin{eqnarray}\label{eq:T_sc,non-rela}
T_{\rm dur,sc} \sim \Delta T \sim 1\,{\rm s}\,(t_{\rm j}+t_{\rm br}+t_{\rm dur})_{0}.
\end{eqnarray}
Note that $T_{\rm dur,sc} \sim t_{\rm dur}$ if $t_{\rm j} \ll t_{\rm br}$ since $t_{\rm br}\lesssim t_{\rm dur}$.
A non-relativistic scatterer scatters photons to a wide angle.
Then, the luminosity of the scattered emission in Eq. (\ref{eq:L_sc,iso}) is
\begin{eqnarray}\label{eq:L_sc,non-rela}
L_{\rm sc} &\sim&3\times10^{46}\,{\rm erg}\,{\rm s}^{-1} \nonumber\\
&\times&\Delta\theta_{-0.5}\Gamma_{\rm j,3}^{-1}\epsilon_{\rm sc,-1} L_{\rm iso,51}
(t_{\rm dur}/T_{\rm dur,sc}).
\end{eqnarray}
The corresponding $\gamma$-ray flux is $\sim4\times10^{-8}$ erg cm$^{-2}$ s$^{-1}$ at the GW detection horizon of LIGO O2, 
$\sim80$ Mpc \citep{2016LRR....19....1A}, which is detectable by {\it Swift}/BAT. 

The time lag for the detection of the scattered emission 
after the GWs in Eq. (\ref{eq:t_lag}) is 
determined by the scatter-starting radius in Eq.~(\ref{eq:radius0})
with Eq.~(\ref{eq:T_scbr}) as
\begin{eqnarray}\label{eq:T_lag,non-rela}
\Delta T_{\rm lag}&\sim&t_{\rm j}+t_{\rm br}+\frac{r_{\rm scbr}}{c\beta_{\rm c}} \nonumber \\
&\sim& 2\,{\rm s}\,(t_{\rm j}+t_{\rm br})_{0}. 
\end{eqnarray}
If $t_{\rm j}\ll t_{\rm br}$, the time lag is comparable to or shorter than the observed duration 
$\Delta T_{\rm lag}\lesssim T_{\rm dur,sc}$ in Eq.~(\ref{eq:T_sc,non-rela}) since $t_{\rm br}\lesssim t_{\rm dur}$. 
If otherwise $t_{\rm j}\gg t_{\rm br}$, the time lag is approximately twice the observed duration
$\Delta T_{\rm lag}\sim 2T_{\rm dur,sc}$. 
In any case, the time lag is comparable to or shorter than twice the observed duration
$\Delta T_{\rm lag}\lesssim 2T_{\rm dur,sc}$. 

\subsection{Relativistic scatterer}\label{sec:rela}

A part of the scatterer could become relativistic.
First the head of the merger ejecta could be relativistic, which could be produced by
the shock breakout at the onset of the binary NS merger \citep{2014MNRAS.437L...6K}.
Second a part of the cocoon could be relativistic due to the incomplete mixing of 
the shocked jet component and the shocked ejecta component 
\citep[e.g., ][]{2017ApJ...834...28N, 2018MNRAS.473..576G, 2017ApJ...848L...6L}.
The mass of the relativistic scatterer is uncertain.
The fraction of the relativistic mass,
which is ejected at the onset of the merger, could be $\lesssim10^{-5}M_{\odot}$ \citep{2014MNRAS.437L...6K}.
The mass of the relativistic cocoon highly depends 
on the degree of the turbulence in the cocoon.
Some hydrodynamic simulations suggest the relativistic cocoon mass of $\sim10^{-7}-10^{-5}M_{\odot}$ 
\citep{2018MNRAS.473..576G, 2017ApJ...848L...6L}.

Because of the relativistic motion of the scatterer, 
even if the jet breakout occurs at $\sim10^{10}$ cm, 
the scattering radius could be far from the breakout point. 
At the scattering radius, the optical depth for the Thomson scattering is approximately
$\tau\sim1$, as discussed in Eqs.~(\ref{eq:tau}) and (\ref{eq:opticaldepth}).
In the relativistic case, the number density in the comoving frame is 
$n'=M_{\rm c}/[4\pi m_{\rm p}r_{\rm sc}^2(r_{\rm sc}/\Gamma_{\rm c}^2)\Gamma_{\rm c}]$, 
so that the scattering radius is
\begin{eqnarray}\label{eq:r_sc,rela}
r_{\rm sc}\sim10^{12}M_{\rm c,-7}^{1/2}~{\rm cm},
\end{eqnarray}
where $M_{\rm c}$ includes only the relativistic component. 

The prompt emission can occur below these scattering radii in Eq.~(\ref{eq:r_emi}).
The radii of internal shocks are around 
$\sim \Gamma_{\rm j}^2 c \Delta t 
\sim 3 \times 10^{11}\,{\rm cm}\,(\Delta t/{\rm ms})
(\Gamma_{\rm j}/10^2)^2$ where $\Delta t$ is the variability timescale 
\citep[e.g., ][]{1994ApJ...427..708P, 1994ApJ...430L..93R, 1997ApJ...490...92K}.
The photospheric radius in Eq.~(\ref{eq:photospheric}) is large 
if $\eta$ and $\Gamma_{\rm j}$ are small.
Poynting-dominated jets generally have large emission radii \citep[e.g., ][]{2005A&A...430....1G, 2015PhR...561....1K} \citep[but see also ][]{2017MNRAS.468.3202B}.

For large scattering radii,
the observed duration of the scattered emission $T_{\rm dur, sc}$ in Eq.~(\ref{eq:T_sc,emi})
is mainly determined by the duration of a single pulse $\Delta T$ in Eq.~(\ref{eq:T_sc}).
If $\Delta T$ is longer than the on-axis sGRB duration $t_{\rm dur}$ which is typically $t_{\rm dur} \sim 0.1\,{\rm s} < 2$ s,
the observed duration is 
\begin{eqnarray}\label{eq:T_sc,rela}
T_{\rm dur,sc} \sim\Delta T \sim 2\,{\rm s}\, r_{\rm sc,12} \Gamma_{\rm c,0.5}^{-2},
\end{eqnarray}
not the engine duration $t_{\rm dur}$.
Note that we assume $\Gamma_c^{-1}>\theta_v-\Delta \theta$ in the above equation.

The luminosity of the scattered emission $L_{\rm sc}$ is obtained from Eq.~(\ref{eq:L_sc,iso}) as
\begin{eqnarray}\label{eq:L_sc,rela}
L_{\rm sc} &\sim& 2\times10^{46}\,{\rm erg}\,{\rm s}^{-1}
\nonumber\\
&\times& \Gamma_{\rm j,2}^{-1} \Delta \theta_{-0.5} 
t_{\rm dur,-1} \Gamma_{\rm c,0.5}^4 r_{\rm sc,12}^{-1} \epsilon_{\rm sc,-2} L_{\rm iso,51}.
\end{eqnarray}
If $\Delta T > t_{\rm dur},~t_{\rm j}$,
the observed time lag in Eq.~(\ref{eq:t_lag}) is also
\begin{eqnarray}\label{eq:T_lag,rela}
\Delta T_{\rm lag}\sim\Delta T_{\rm scbr}\sim  2\,{\rm s}\, r_{\rm scbr,12} \Gamma_{\rm c,0.5}^{-2},
\end{eqnarray}
since $t_{\rm br} \lesssim t_{\rm dur}$,
and is comparable to the duration $\Delta T_{\rm lag} \sim T_{\rm dur, sc}$
in Eq.~(\ref{eq:T_sc,rela}).

\subsection{Spectrum}\label{sec:spec}

\begin{figure*}
 \begin{center}
  \includegraphics[width=120mm]{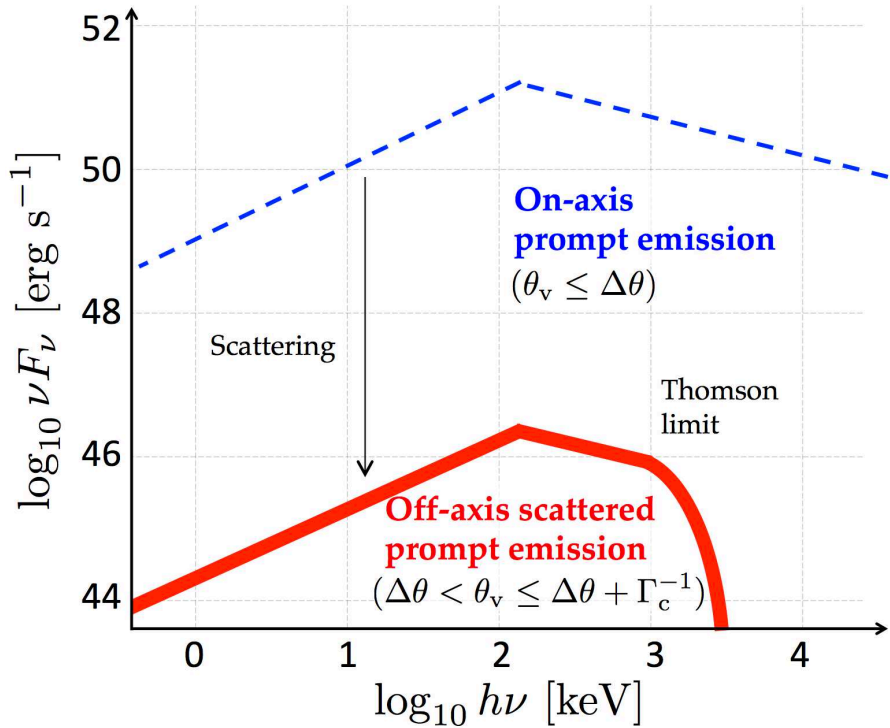}
   \caption{The energy spectrum of the prompt and its scattered emission from an sGRB.}
  \label{fig:spectrum}
 \end{center}
\end{figure*}

Figure \ref{fig:spectrum} shows the energy spectrum 
of the scattered prompt emission ({\it red solid curve}).
The basic features are the same for both the non-relativistic and relativistic cases.
For the non-relativistic scatterer,
the electrons see the same spectrum that the on-axis observers see.
The Thomson scattering copies the original spectral shape 
below $\sim m_e c^2 \sim {\rm MeV}$ energy range.
Above this energy, the scattered spectrum has a cutoff
because the Klein-Nishina effect reduces the cross section
and makes the scattering angles anisotropic.
For the relativistic scatterer, 
the electrons see the red-shifted spectrum in the comoving frame, 
while the scattered photons are blue-shifted by the bulk motion of the scatterer in the observer frame. 
Since $\Gamma_{\rm j}\gg\Gamma_{\rm c}$, the energy of the scattered photon is 
\begin{eqnarray}
E_{\rm sc}=\frac{E}{2}\left[1+\frac{\beta_{\rm c}(\mu_{\rm sc}-\beta_{\rm c})}{1-\beta_{\rm c}\mu_{\rm sc}}\right],
\end{eqnarray}
where $E$ is the energy of the incident photon, 
and $\theta_{\rm sc}=\cos^{-1}\mu_{\rm sc}$ is the angle of the scattered photon direction
relative to the motion of the scatterer. 
For $\theta_{\rm sc}\le1/\Gamma_{\rm c}$, 
the energy of the photon changes to $E_{\rm sc}/E\sim0.5-1$ by a scattering, 
resulting in a similar spectral shape to the non-relativistic case. 

The peak energy $E_{\rm p}$ is similar to those of on-axis sGRBs.
This is the characteristic of the scattering model,
which is the main difference from 
other models such as the off-axis de-beamed emission and 
the emission from low-$\Gamma_{\rm j}$ jet/cocoon components 
as discussed in Sec. \ref{sec:discussion}.

For on-axis sGRBs, there are an $E_{\rm p}$--$L_{\rm p}$ correlation
between the peak energy $E_{\rm p}$ and the peak luminosity $L_{\rm p}$ \citep{2004ApJ...609..935Y, 2013MNRAS.431.1398T},
and also an $E_{\rm p}$--$E_{\rm iso}$ correlation \citep{2002A&A...390...81A, 2013MNRAS.431.1398T}.
The off-axis scattered sGRBs appear below these correlations 
in the $E_{\rm p}$--$L_{\rm p}$ and $E_{\rm p}$--$E_{\rm iso}$ diagrams.
Since $L_{\rm sc} \propto L_{\rm iso}$ in Eqs.~(\ref{eq:L_sc,non-rela}) and (\ref{eq:L_sc,rela}), 
we may expect similar correlations like
$E_{\rm p}$--$L_{\rm sc}$, probably
with a larger dispersion caused by the dependence on many other parameters.

\subsection{Monte Carlo Simulation}
\label{sec:MC}

In order to estimate the isotropic energy of the scattered component in more realistic conditions, 
we perform Monte Carlo simulation for the photon propagation. 
Especially, there is no clear boundary between the jet and ejecta in reality. 
Some numerical simulations show that the energy and Lorentz factor are seen 
to smoothly vary from the jet to the cocoon 
\citep[e.g., ][]{2003ApJ...586..356Z, 2005A&A...436..273A, 2014ApJ...784L..28N, 2014ApJ...788L...8M, 
2017arXiv171203237L, 2017ApJ...835L..34M}. 
Then, we assume an ultra-relativistic core with a uniform emissivity surrounded by the ejecta with mildly relativistic velocity. 
As a simple toy model, we linearly interpolate the difference of the Lorentz factor between 
the jet and the cocoon as
\begin{eqnarray}\label{eq:Gamma(theta)}
\Gamma(\theta) = \Gamma_{\rm j}-(\Gamma_{\rm j}-\Gamma_{\rm c})\frac{\theta-\Delta\theta}{\theta_{\rm int}}
~~(0\le\theta-\Delta\theta\le\theta_{\rm int}),
\end{eqnarray}
where $\theta$ is the polar angle in a spherical polar coordinate. 
Here, we introduce the angle of the interpolating region $\theta_{\rm int}$ as a model parameter. 
Similar interpolation is also assumed for the electron number density. 
Using this model, we consider some smooth distributions for the velocity and the optical depth, 
and investigate the effects on the isotropic energy of the scattered component. 

The parameters of our model are summarized in Table 1. 
We assume that the jet and ejecta are steady radial flows 
with the Lorentz factor of $\Gamma(\theta)$ given by Eq.~(\ref{eq:Gamma(theta)}), 
and the thermal motion of electrons in the flows is negligible. 
Photons are emitted to an isotropic direction in the jet comoving frame at the emission site.  
The emission site is at a given radius of $r_{\rm emi}$ and polar angle of $\theta\le\Delta\theta$. 
The total number of the generated photons in each simulation is about $10^9$. 
We calculate the scattering probability $P$ for each photon, 
\begin{eqnarray}
P=1-\exp(-\tau).
\end{eqnarray}
The optical depth $\tau$ for each photon is derived by the integration along the photon path
from the emission point. 
We generate a random number in the range of [0:1] for each photon.
If the scattering probability reaches the given random number, the photon is scattered. 
For the scattering process, we consider Thomson scattering. 
The probability of the photon scattered into each solid angle follows 
the differential cross section of Thomson scattering \citep[e.g., ][]{1979rpa..book.....R} 
in the comoving frame. 
We generate additional two random numbers to determine the photon direction after the scattering \citep{1983ASPRv...2..189P}.
Then, we generate a new random number for a scattered photon, and calculate the scattering probability
from the scattering point.
Because of the computational cost, we do not take into account the contribution of more than six-time scattered photons
to the total radiation energy.
It is noted that the once and twice scattered photons dominantly contribute to the total number of escaping photons
with the off-axis direction.
Escaping photons are counted at the surface of a sphere with a radius of $100~r_{\rm emi}$
as a function of the propagation direction. 

\begin{figure}
 \begin{center}
  \includegraphics[width=60mm, angle=270]{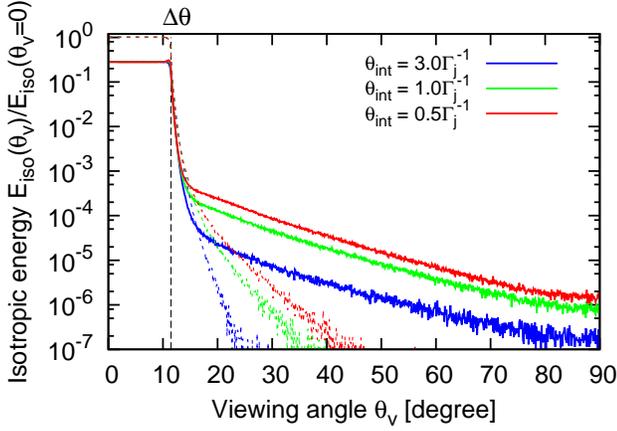}
   \caption{The isotropic photon energies of the scattered (thick solid curves) and unscattered components (thin dashed curves) normalized by that of unscattered component with $\theta_{\rm v}=0$
 as a function of the viewing angle $\theta_{\rm v}$.}
  \label{fig:montecarlo}
 \end{center}
\end{figure}

Figure \ref{fig:montecarlo} shows the photon isotropic energy as a function of the viewing angle. 
The vertical values are normalized by the isotropic energy of the unscattered component with $\theta_{\rm v}=0^\circ$. 
The normalized isotropic energy of the scattered component (thick solid curves) is $\sim10^{-3}-10^{-4}$, 
for the case of $\theta_{\rm int}\lesssim\Gamma_{\rm j}^{-1}$ and the viewing angle of $\theta_{\rm v}\sim25^\circ-30^\circ$, 
which gives the duration and time delay $\sim 2$ sec (Eqs.~\ref{eq:T_sc,rela} and \ref{eq:T_lag,rela}). 
The photons scattered by the ejecta with low Lorentz factor contribute to the radiation energy with a large viewing angle.
Since the number of photons emitted from the jet decreases with an angle $\theta$, 
the number of photons scattered by the ejecta increases for the narrower interface of  $\theta_{\rm int}$.
The isotropic energy is comparable to the analytical result of $L_{\rm iso}\Delta T_{\rm dur,sc}$ 
in Eqs.~(\ref{eq:T_sc,rela}) and (\ref{eq:L_sc,rela}). 
For the larger angle of $\theta_{\rm v}\gtrsim30^\circ$, isotropic energy is reduced 
because of the beaming effect (Eq.~\ref{eq:opening}) 
and the enhanced optical depth for large angle photons relative to the flow direction. 
We also show the isotropic energy of the unscattered component in Figure \ref{fig:montecarlo} (thin dashed curves). 
For the angle of $\theta_{\rm v}\gtrsim15^\circ$, the scattered component dominantly contributes to the isotropic photon energy. 
Although the isotropic energy of the scattered component with a range of direction of $\theta_{\rm v}\gtrsim15^\circ$ is 
$E_{\rm iso}(\theta_{\rm v})/E_{\rm iso}(0)\lesssim10^{-4}$ in the case of $\theta_{\rm int}\gtrsim\Gamma_{\rm j}^{-1}$, 
the detail analysis including the dependence of other model parameters will be presented in a forthcoming paper.

\begin{table}
 \begin{center}
  \begin{tabular}{llc}
\multicolumn{3}{c}{TABLE 1 Model Parameters} \\ \hline
Symbol & & Value \\ \hline
$r_{\rm emi}$ & Photon emission radius & $10^{12}$ cm        \\
$\Delta\theta$  & Jet opening angle & $0.2$ rad   \\
$\Gamma_{\rm j}$ & Jet Lorentz factor & $200$   \\
$L_{\rm iso}$ & Jet isotropic luminosity & $10^{51}$ erg s$^{-1}$   \\
$M_{\rm c}$ & Cocoon mass & $5\times10^{-8}M_{\odot}$ \\
$\Gamma_{\rm c}$ & Ejecta Lorentz factor & $3$ \\
$\theta_{\rm int}$ & Angle between jet and ejecta & $0.5\Gamma_{\rm j}^{-1}, 1.0\Gamma_{\rm j}^{-1}, 3.0\Gamma_{\rm j}^{-1}$ \\ \hline
 \label{tab:parameter}
  \end{tabular}
 \end{center}
 \end{table}

\section{Implications of sGRB 170817A}
\label{sec:implications}

Now let us consider the implications of sGRB 170817A to the scattering model. 
In sGRB 170817A, the duration, time-averaged luminosity, and time lag
of the prompt emission are
$T_{\rm dur,sc}\sim2$ s, $L_{\rm sc}\sim2.5\times10^{46}$ erg s$^{-1}$, and $\Delta T_{\rm lag}\sim1.7$ s, 
respectively \citep{2017arXiv171005446G}.
The estimated observed angle with respect to the orbital axis of the pre-merging binary system, 
which is supposed to be equal to the jet axis, is 
$\theta_{\rm v} \lesssim 32 \ \rm deg$ \citep{2017PhRvL.119p1101A, 2017Natur.551...85A}.  

In the non-relativistic case, 
from Eqs.~(\ref{eq:T_sc,non-rela}) and (\ref{eq:T_lag,non-rela}),
the observed duration $T_{\rm dur,sc}$ 
and time lag $\Delta T_{\rm lag}$ are explained if
the engine duration and the breakout timescale are
$t_{\rm dur}\sim t_{\rm br}\sim2$ s $(> t_{\rm j})$
or the jet-launch time is $t_{\rm j} \sim 2$ s $(> t_{\rm dur}, t_{\rm br})$.
From Eqs.~(\ref{eq:radius}) and (\ref{eq:photospheric}) and 
the condition $r_{\rm emi}\lesssim r_{\rm sc}$ in Eq.~(\ref{eq:r_emi}), 
the Lorentz factor of the jet should satisfy the condition $\eta\ge\Gamma_{\rm j}\gtrsim500$.
Then, the observed luminosity is consistent with the scattering model 
within the reasonable ranges of the parameters:
$\Gamma_{\rm j}\sim10^3$,
$\Delta\theta\sim 0.1$--$0.3$ rad, $\epsilon_{\rm sc}\sim0.1$, and $L_{\rm iso}\sim10^{51}$ erg s$^{-1}$
in Eq.~(\ref{eq:L_sc,non-rela}). 
For this parameter set, 
the emission and scattering sites are $r_{\rm sc}\sim r_{\rm emi}\sim10^{10}$ cm and the photon-to-baryon-ratio is $\eta\sim 10^3$, 
which would give constraints on the models of the prompt emission \citep[e.g., ][]{2000ApJ...530..292M, 2011PThPh.126..555I}.

In the relativistic case, 
the observed duration and time lag can be explained if
the scattering radius is $r_{\rm sc}\sim10^{12} (\Gamma_{\rm c}/3)^{2}$ cm
from Eqs.~(\ref{eq:T_sc,rela}) and (\ref{eq:T_lag,rela}).
If the Lorentz factor of the relativistic scatterer is
$\Gamma_{\rm c} \sim 1$--$10$,
the observed luminosity is also explained by the reasonable ranges of the parameters:
$\Gamma_{\rm j}\sim10^2$,
$\Delta\theta\sim 0.1$--$0.3$ rad,
$t_{\rm dur} \sim 0.1$ s,
$\epsilon_{\rm sc}\sim0.01$, and $L_{\rm iso}\sim10^{51}$ erg s$^{-1}$
in Eq.~(\ref{eq:L_sc,rela}).
From Eq.~(\ref{eq:r_sc,rela}) with $\Gamma_{\rm c} \sim 1$--$10$,
the mass of the relativistic material should be 
$M_{\rm c}\sim 10^{-8}$--$10^{-6} M_{\odot}$.
Note that the opening angle of the scattered emission in Eq.~(\ref{eq:opening})
is $\Delta\theta_{\rm sc} \sim 0.3 (\Gamma_{\rm c}/3)^{-1}$ rad,
which is small for large $\Gamma_{\rm c}$,
and hence too large $\Gamma_{\rm c}$ is not preferred.

The observed peak energy of the main pulse 
$E_{\rm p}\sim 185 \pm 62$ keV
is consistent with the scattering model
because previous sGRBs that are likely on-axis have comparable peak energies.
In particular, the $E_{\rm p}$--$L_{\rm p}$ correlation suggests $L_{\rm p}\sim2\times10^{51}$ erg s$^{-1}$ 
for the main peak \citep{2013MNRAS.431.1398T}, which is consistent with our choice of $L_{\rm iso}$ in the above discussion.
The temperature of the weak tail $kT=10.3 \pm 1.5$ keV
is not consistent with the scattering model,
while the off-axis de-beamed emission can explain such low peak energies
\citep{2017arXiv171005905I}.
Both the scattering model and the off-axis model coexist in a single sGRB,
for example if the Lorentz factor of the jet is high in the main pulse
and low in the weak tail.
The emission region could also range from
the photosphere (for scattering) to
the internal shocks (for off-axis de-beaming).

\section{On jet-cocoon Structure}
\label{sec:jet-cocoon}

\begin{figure}
 \begin{center}
  \includegraphics[width=60mm, angle=270]{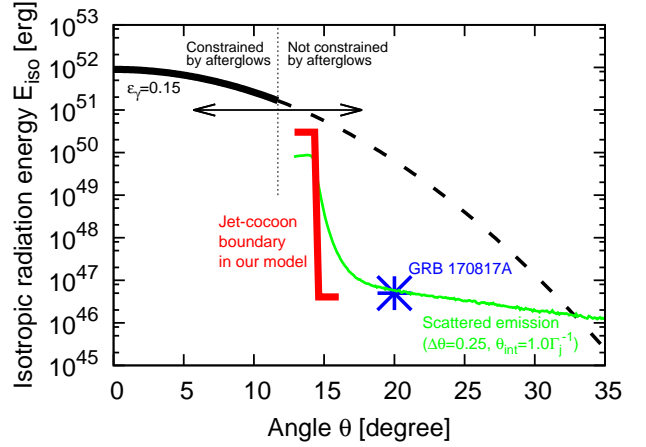}
  \caption{The distributions of isotropic radiation energy for the structured jet model
    \citep[black curve;][]{2018arXiv180502870A,2018ApJ...856L..18M,2018arXiv180409345X}, 
    and our model with $\theta_{\rm int}=\Gamma_{\rm j}^{-1}$ (red curve)
    assuming the radiation efficiency $\epsilon_{\gamma}=0.15$.
    The scattered emission in our model (green curves) is the same except for the jet opening angle of $\Delta\theta=0.25$ rad.
    The isotropic energy of sGRB 170817A is shown as a blue asterisk. 
    Note that the cocoon part of the red line assumes $\epsilon_{\gamma}=0.15$ here but the actual radiation 
    efficiency would be much fainter.
  }
  \label{fig:distribution}
 \end{center}
\end{figure}

The energy and the Lorentz factor profiles of the jet and cocoon at the prompt emission phase are highly uncertain. 
In this circumstance, we should be open-minded about possible jet structures that are consistent with observations and theories. 
In section \ref{sec:MC}, we consider the idealized structure between the jet and cocoon by using the parameter of $\theta_{\rm int}$. 
This prescription is very useful for future studies. 
We show that if the jet has relatively steep angular distributions of the energy and Lorentz factor ($\theta_{\rm int}\sim\Gamma_{\rm j}^{-1}$),
the properties of sGRB 170817A could be explained by the scattering emission of the canonical sGRB.
On the other hand, if we adopt a smoothly varying jet ($\theta_{\rm int}\gg\Gamma_{\rm j}^{-1}$),
most photons cannot reach to the low-$\Gamma$ component ($\Gamma<10$). 
Then, the radiation energy of the scattering emission for an off-axis observer would become much lower as shown in Figure \ref{fig:montecarlo}
with $\theta_{\rm int}=3\Gamma_{\rm j}^{-1}$. 
In this case, it is difficult to explain the observed properties of sGRB 170817A based on the scattering model.
In the following, we discuss the jet-cocoon structure suggested by the numerical simulation results and the afterglow observations. 

Some numerical simulations show that after the jet penetrates the ejecta,
the width of $\theta_{\rm int}$ is much wider than the inverse of Lorentz factor of the ultra-relativistic component,
$\Gamma_{\rm j}^{-1}$ \citep[e.g., ][]{2017arXiv171203237L}. 
However, it is difficult to calculate the steep angular distributions of the energy and Lorentz factor by current numerical simulations
due to the following numerical effects. 
First, in order to resolve the steep structure, an angular resolution of $d\theta\lesssim\Gamma_{\rm j}^{-1}\sim0.005$ rad is required. 
The 3D numerical simulation is performed to avoid the plug effect, 
which is considered as a numerical artifact of the symmetry
\citep{2018MNRAS.473..576G}. 
Typical angular resolution at the base of the jet in 3D simulation is $d\theta\gtrsim0.1$ rad,
which is not enough to resolve the structure in our model. 
Second, artificial baryon loading could happen when the contact discontinuity
between the jet and cocoon crosses the grids via numerical diffusion \citep{2013ApJ...777..162M}.
A certain level of baryon loading is unavoidable, and could also smooth the angular distributions. 

In addition, there are some uncertainties for the set-up of numerical simulations such as the 3D structures
of the ejecta and the magnetic field, and the energy injection manner from the jet to the ejecta.
For example, if the ejecta density is low enough, 
the jet can keep the initial structure, leading to a sharp boundary. 
Future more sophisticated numerical simulations are required to clarify
whether the steep angular distributions considered in our model are realized or not.

The afterglow emission has been detected in sGRB 170817A, which could give constraints on the jet structure. 
\citep{2017Natur.551...71T, 2017Sci...358.1565E, 2017Sci...358.1579H, 2017arXiv171008514F, 2017ApJ...848L..20M,
  2017ApJ...848L..21A, 2017ApJ...848L..25H, 2017ApJ...850L..21K}.
The observed light curves show a gradual rise in radio, optical, and X-ray bands up to $\sim$ 150 days
\citep{2018ApJ...859L..23P,2018ApJ...853L...4R,2018Natur.554..207M,2018arXiv180102669L,2018ApJ...856L..18M,2018arXiv180302768R,
  2018ApJ...858L..15D,2018arXiv180502870A,2018arXiv180504093N,2018MNRAS.478L..18T,2018A&A...613L...1D,2018arXiv180609693M,2018arXiv180800469G}. 
The gradual rise implies the continuous energy injection into the observed region,
which is consistent with the structured jet with smoothly varying kinetic energy distribution in angular direction. 

However, the afterglow observations cannot in principle constrain the Lorentz factor distribution
at the prompt emission phase, because the afterglow emission comes from the decelerated shocks and
does not keep the information about the initial Lorentz factor of the jet and cocoon  
\citep[e.g., ][]{1997ApJ...489L..37S}. 
Even if the kinetic energy of the jet has a smooth angular distribution, 
the scattering could occur if there is a steep distribution in the Lorentz factor. 

For the energy distribution,
the angular structure may be different between the prompt emission phase and the afterglow emission phase.
In about a half of sGRBs, the isotropic photon energy of the prompt emission is lower than that of the later activities,
so-called extended emission \citep{2017ApJ...846..142K}.
The injected energy from the jet component, which causes the extended emission, may dominate the observed afterglow emission. 
In this case, the afterglow emission does not constrain the jet structure at the prompt emission phase.

Even if the jet component of the prompt emission is the dominant energy source of the observed afterglow emission,
the structures may be different between the afterglow phase and prompt emission phase. 
During the propagation in the circum-merger medium, the energy distribution becomes smooth even if the initial jet
has a sharp profile because the material at the boundary of the jet expands to a wider angle.
This is actually observed in the numerical simulation of afterglows 
\citep{2001grba.conf..312G}. 
As in Fig. 2 of \citet{2002ApJ...570L..61G}, the numerical calculations show gradual rising light curves of afterglow
even from a top-hat jet, which is caused by the side expansion of the material from the jet.  
Therefore, the afterglow observation cannot rule out the existence of
the jet with steep angular distributions at the prompt emission phase.

Furthermore, as shown in Fig. \ref{fig:distribution},
the energy distribution in our model is consistent with and even implied by the observations. 
We compare the isotropic radiation energy of the prompt emission from the structured jet 
constrained by the afterglow observations with the observed value of sGRB 170817A. 
Here, we assume that the angular structure at the afterglow emission phase is 
the same as that at the prompt emission phase. 
We use the radiation efficiency of $\epsilon_{\gamma}$ and 
the isotropic kinetic energy distribution of the structured jet
to estimate the isotropic radiation energy of the prompt emission. 
We adopt the structured jet model suggested by \citet{2018arXiv180502870A} 
\citep[see also ][]{2018ApJ...856L..18M,2018arXiv180409345X},
the radiation efficiency of $\epsilon_{\gamma}=0.15$, the circum-merger density
$n_{\rm ism}\sim10^{-4}$ cm$^{-3}$, and the viewing angle of $\theta_{\rm v}\sim20^{\circ}$,
which are consistent with the constraints by the observed afterglow emission, 
including the recent VLBI observations \citep{2018arXiv180609693M,2018arXiv180800469G}.
However, as seen in Figure \ref{fig:distribution}, 
a simple extrapolation of the jet structure (black dashed curve) to the observed 
viewing angle exceeds the observed one in sGRB 170817A (blue asterisk)
unless the radiation efficiency of $\epsilon_{\gamma}$ is much smaller than that 
in typical prompt emission of GRBs \citep[e.g., ][]{2001ApJ...547..922F,2016MNRAS.461...51B}.
This suggests a steeper structure than a simple extrapolation in the outer part of the jet
at the prompt emission phase. 
In fact, the afterglow observations cannot give constraints on the structure around the viewing angle, 
in particular $|\theta_{\rm v}-\theta|\lesssim1/\Gamma\sim10^\circ$, 
because this part contributes to the earlier emission and 
does not affect the emission after the first detection of the afterglow, 
i.e., $>$ 9 days after the merger
\citep[e.g., ][]{2017Natur.551...71T}. 
A steeper energy distribution such as our model (red curve) and narrow Gaussian jet models
\citep{2018ApJ...853L...4R,2018arXiv180102669L,2018MNRAS.478L..18T}
is implied for the outer region of the jet 
\footnote{We should note that the cocoon emission would be much fainter than the outer red line in Figure \ref{fig:distribution}
  because a typical radiation efficiency of a cocoon is much smaller than our fiducial value of $\epsilon_{\gamma}=0.15$.}. 
The detection of the earlier afterglow emission is important to clarify the detailed structure of the jet and cocoon. 

\section{Summary and DISCUSSIONS} 
\label{sec:discussion}

We propose that the Thomson-scattered emission of a prompt sGRB 
is a promising EM counterpart to a binary NS merger,
which is observable from a large viewing angle.
The GRB jet is surrounded by a cocoon, either sub-relativistic or mildly-relativistic,
which could scatter the prompt emission into a wide angle 
$\sim 1/\Gamma_{\rm c}$.
The striking feature of the scattering model is that 
the spectral peak energy $E_{\rm p}$ 
is almost independent of the viewing angle
and similar to the on-axis values if $E_{\rm p} \lesssim {\rm MeV}$ 
as in Fig.~\ref{fig:spectrum}.
The isotropic-equivalent luminosity of the scattered emission
is also estimated as $L_{\rm sc}\sim10^{46}-10^{47}$ erg s$^{-1}$ 
in Eqs.~(\ref{eq:L_sc,non-rela}) and (\ref{eq:L_sc,rela})
for fiducial parameters
if the prompt emission occurs around the scattering radius.
These features are consistent with sGRB 170817A, 
in particular the main pulse of the prompt emission,
supporting the scattering model.
We also clarify what determine
the duration of the scattered emission and its time lag behind the GWs
in Eqs.~(\ref{eq:T_sc,non-rela}) and (\ref{eq:T_lag,non-rela})
for the non-relativistic case
and in Eqs.~(\ref{eq:T_sc,rela}) and (\ref{eq:T_lag,rela})
for the relativistic case.
These estimates also reproduce
the observed duration $\sim 2$ s and time lag $\sim 1.7$ s
in sGRB 170817A with typical sGRB parameters. 
We also perform the Monte Carlo simulation, and find that 
the luminosity of the scattering component is consistent with the observations. 
The scattering model requires the steep angular distributions of 
the Lorentz factor structure between the jet and cocoon, 
which is consistent with the late-time afterglow observations 
(see discussion in section \ref{sec:jet-cocoon}).  
These successful results support that 
the binary NS merger in GW170817 is associated with a typical sGRB jet,
which produces sGRB 170817A and other EM counterparts.
They also suggest that the scattering emission is ubiquitous in off-axis sGRBs
and provides a clear target for future simultaneous observations with GWs.

We should remind that the scattering emission coexists 
with the off-axis de-beamed emission,
complementing each other.
In particular, both the models suggest a typical off-axis sGRB jet for sGRB 170817A.
The possible evolution of the observed peak energies
from the main pulse ($E_{\rm p}\sim 185 \pm 62$ keV) 
to the weak tail ($kT\sim 10.3\pm 1.5$ keV)
may be explained by invoking the scattering and the off-axis de-beamed emissions, respectively.

The remarkable characteristic of the scattering model is that 
the spectral shape is similar to the on-axis one with a cutoff around $\sim {\rm MeV}$ energy
due to the Klein-Nishina effect as in Fig.~\ref{fig:spectrum}.
In particular, the peak energy $E_{\rm p}$ distribution is similar to the on-axis one
with an accumulated bump around MeV energy.
We also expect $E_{\rm p}$--$L_{\rm sc}$ and $E_{\rm p}$--$E_{\rm iso}$
correlations with similar slopes, fainter normalizations, 
but larger dispersions than the on-axis correlations.
These spectral features are the main difference from the other models 
such as the off-axis de-beamed emission from an sGRB jet \citep[e.g., ][]{2017arXiv171005905I}
and the emission from a low-$\Gamma_{\rm j}$ jet/cocoon with a wide opening angle
\citep[e.g., ][]{2017MNRAS.472.4953L, 2018MNRAS.473L.121K, 2017MNRAS.471.1652L, 2017ApJ...848L...6L}.
The off-axis jet model generally predicts low peak energies $\sim10$ keV \citep[e.g.,][]{2001ApJ...554L.163I, 2017arXiv171005905I}.
In the low-$\Gamma_{\rm j}$ jet and mildly-relativistic cocoon models, 
if the emission is purely thermal radiation, 
the peak energy is also significantly soft ($\sim1-10$ keV) compared with the on-axis prompt emission \citep{2017ApJ...848L...6L}.
Therefore, we can test models by observing the peak energy distributions.
The other observables such as the duration and time lag after GWs
could be also helpful for discriminating models.

The scattering layer could be accelerated by the illumination of the prompt emission.
If the kinetic energy of the scatterer in the illuminated region is less than
the energy injected by the radiation,
the scattering layer is accelerated. 
The radiation energy which illuminates the ejecta is
\begin{eqnarray}
E_{\rm rad}&\sim&\frac{2}{\Delta\theta\Gamma_{\rm j}}\frac{\Delta\theta^2}{2}E_{\rm iso}\epsilon_{\rm acc} \nonumber \\
&\sim&3\times10^{46}\Delta\theta_{-0.5}\Gamma_{\rm j,2}^{-1}\epsilon_{\rm acc,-1}E_{\rm iso,50}~{\rm erg},
\end{eqnarray}
where $\epsilon_{\rm acc}$ is the energy gain normalized
by the electron rest mass energy for single Thomson scattering, 
which is $(E_{\rm p}/m_{\rm e}c^2)^2\sim0.1$ for photons with $E_{\rm p}\sim200$ keV.
In the non-relativistic scatterer case, the kinetic energy of the illuminated ejecta is 
\begin{eqnarray}
E_{\rm kin}&\sim&\frac{1}{2}f_{\rm cm}M_{\rm e}\beta_{\rm c}^2c^2\frac{3\Delta\theta}{4\pi\Gamma_{\rm j}} \nonumber \\
&\sim&4\times10^{47}\Delta\theta_{-0.5}\Gamma_{\rm j,2}^{-1}f_{\rm cm,-0.3}M_{\rm e,-2}\beta_{\rm c,-0.5}^2~{\rm erg}, 
\end{eqnarray}
so that the acceleration by the radiation is negligible.
On the other hand, in the relativistic scatterer case,
the kinetic energy of the illuminated ejecta is 
\begin{eqnarray}
E_{\rm kin}&\sim&\Gamma_{\rm c}M_{\rm c}c^2\frac{3\Delta\theta}{4\pi\Gamma_{\rm j}} \nonumber \\
&\sim&4\times10^{44}\Delta\theta_{-0.5}\Gamma_{\rm j,2}^{-1}\Gamma_{\rm c,0.5}M_{\rm c,-7}~{\rm erg}, 
\end{eqnarray}
which is less than the radiation energy, $E_{\rm rad}$. 
Then, the scattered layer could be accelerated. 
During the acceleration, the illuminated ejecta interact with a part of the outer region of ejecta
($\theta>\Delta\theta+1/\Gamma_{\rm j}$).
Then, a part of the injected energy would be distributed.
We assume that the ejecta with an angle of $\Gamma_{\rm acc}^{-1}$ are accelerated
up to $\Gamma_{\rm acc}$ as a result of the interaction with the outer region of the ejecta,
\begin{eqnarray}
E_{\rm rad}=\Gamma_{\rm acc}\frac{M_{\rm c}c^2}{2\Gamma_{\rm acc}^2},
\end{eqnarray}
where $\sim 1/(2\Gamma_{\rm acc}^2)$ is the volume fraction of the accelerated ejecta. 
Then, the Lorentz factor of the accelerated ejecta is 
\begin{eqnarray}
\Gamma_{\rm acc}&\sim&\frac{M_{\rm c}c^2}{2E_{\rm rad}} \nonumber \\
&\sim&3\Delta\theta_{-0.5}^{-1}\Gamma_{\rm j,2}\epsilon_{\rm acc,-1}^{-1}M_{\rm c,-7}E_{\rm iso,50}^{-1},
\end{eqnarray}
which is similar to the pre-accelerated value as we consider in section \ref{sec:scatterer}.
In this case, the radiative acceleration does not significantly change the results in section \ref{sec:scatterer}.
The time-dependent radiation-hydrodynamics simulations are required to determine
the detail profile of the accelerated ejecta.
This will be addressed in future works.

The multi-scattering components could also contribute to the soft weak tail.
If photons are trapped by the expanding cocoon via scattering, 
the adiabatic cooling decreases the photon temperature or peak energy 
while keeping the photon number.
This is an interesting future problem. 

The success of the scattering model has also
interesting implications for the emission mechanism of sGRBs.
For the non-relativistic scatterer, 
the emission and scattering sites have to be $\lesssim10^{10}$ cm to scatter the prompt emission 
in Eqs.~(\ref{eq:r_emi}) and (\ref{eq:radius}).
Based on the photosphere model, the Lorentz factor and the photon-to-baryon ratio should be 
$\Gamma_{\rm j}\sim 10^3$ and $\eta\sim 10^3$ from Eq.~(\ref{eq:photospheric})
and discussions in Sec.~\ref{sec:implications}.
For the relativistic scatterer, the scattering site is preferred not to be far away
$r_{\rm sc}\lesssim10^{12}$ cm
from Eq. (\ref{eq:r_sc,rela}) and discussions in Sec.~\ref{sec:implications}.
In either case, the emission radius is not very large
as in the case of some Poynting-dominated models,
favoring the photosphere model and the internal shock model.

Although scattered emission may have been already detected in the previous observations,
they would be most likely classified as redshift-unknown events,
because they are faint and
their early afterglows are also faint due to the off-axis viewing angle.
The afterglow emission of the scatterer 
is also faint before its deceleration time,
which is late because of its low Lorentz factor.

The event rate of the scattered prompt emission 
detected by a $\gamma$-ray detector is estimated as 
\begin{eqnarray}\label{eq:rate}
{\cal R}&\sim&0.3{\cal R}_{\rm merge,3.5}F_{\rm lim,-7}^{-3/2}L_{\rm sc,46}^{3/2}\Delta\Omega_{4\pi}~{\rm yr}^{-1},
\end{eqnarray}
where we assume isotropic emission,
${\cal R}_{\rm merge,3.5}\equiv{\cal R}_{\rm merge}/10^{3.5}$ Gpc$^{-3}$ yr$^{-1}$ 
is the merger rate of the binary NS, 
$F_{\rm lim}$ is the detector sensitivity limit in $1$ s integration time, 
and $\Delta\Omega_{4\pi}\equiv\Delta\Omega/(4\pi~{\rm steradian})$ is the detector field-of-view. 
Using the upper limit on the binary NS merger rate derived from the LIGO O1
\citep[$\sim3\times10^{3}$ yr$^{-1}$ Gpc$^{-3}$; ][]{2016ApJ...832L..21A}, 
$\sim3$ events would be detected during $\sim10$ yr observations of {\it Fermi}/GBM with a full sky field of view 
and the flux limit $F_{\rm lim}\sim10^{-7}$ erg cm$^{-2}$ \citep{2016ApJS..223...28N}. 
For {\it Swift}/BAT with
the field of view $\Delta\Omega=1.4$ steradian 
and the flux limit $F_{\rm lim}\sim10^{-8}$ erg cm$^{-2}$ \citep{2013ApJS..209...14K}, 
$\sim12$ events would be detected during the $\sim 13$ yr observations. 
The simultaneous observation with GWs will help us not to miss the event to follow up,
revealing the off-axis nature of the sGRBs and binary NS mergers.

\acknowledgments

We are grateful to the anonymous referee for constructive comments.
We would like to thank Kuniaki Masai, Yutaka Ohira, 
Masaru Shibata, Masaomi Tanaka, Shuta Tanaka, Ryo Yamazaki, and Michitoshi Yoshida for discussions.
This work is partly supported by
``New Developments in Astrophysics Through Multi-Messenger Observations
of Gravitational Wave Sources'', No.~24103006 (KI, TN), 
KAKENHI Nos. 16J06773, 18H01246 (SK), Nos. 26287051, 26247042, 17H01126, 17H06131, 17H06362, 17H06357 (KI),
No. 17K14248 (KK), No.~15H02087 (TN)
by the Grant-in-Aid from the Ministry of Education, Culture, Sports,
Science and Technology (MEXT) of Japan.

\bibliographystyle{apj_8}
\bibliography{ref}

\begin{thebibliography}{135}
\expandafter\ifx\csname natexlab\endcsname\relax\def\natexlab#1{#1}\fi

\bibitem[{{Abbott} {et~al.}(2016{\natexlab{a}}){Abbott}, {Abbott}, {Abbott},
  {Abernathy}, {Acernese}, {Ackley}, {Adams}, {Adams}, {Addesso}, {Adhikari},
  \& et~al.}]{2016LRR....19....1A}
{Abbott}, B.~P. {et al.}\  2016{\natexlab{a}}, Living Reviews in Relativity,
  19, 1

\bibitem[{{Abbott} {et~al.}(2016{\natexlab{b}}){Abbott}, {Abbott}, {Abbott},
  {Abernathy}, {Acernese}, {Ackley}, {Adams}, {Adams}, {Addesso}, {Adhikari},
  \& et~al.}]{2016ApJ...832L..21A}
--- 2016{\natexlab{b}}, \apjl, 832, L21

\bibitem[{{Abbott} {et~al.}(2017{\natexlab{a}}){Abbott}, {Abbott}, {Abbott},
  {Acernese}, {Ackley}, {Adams}, {Adams}, {Addesso}, {Adhikari}, {Adya}, \&
  et~al.}]{2017Natur.551...85A}
--- 2017{\natexlab{a}}, \nat, 551, 85

\bibitem[{{Abbott} {et~al.}(2017{\natexlab{b}}){Abbott}, {Abbott}, {Abbott},
  {Acernese}, {Ackley}, {Adams}, {Adams}, {Addesso}, {Adhikari}, {Adya}, \&
  et~al.}]{2017ApJ...848L..13A}
--- 2017{\natexlab{b}}, \apjl, 848, L13

\bibitem[{{Abbott} {et~al.}(2017{\natexlab{c}}){Abbott}, {Abbott}, {Abbott},
  {Acernese}, {Ackley}, {Adams}, {Adams}, {Addesso}, {Adhikari}, {Adya}, \&
  et~al.}]{2017PhRvL.119p1101A}
--- 2017{\natexlab{c}}, Physical Review Letters, 119, 161101

\bibitem[{{Abbott} {et~al.}(2017{\natexlab{d}}){Abbott}, {Abbott}, {Abbott},
  {Acernese}, {Ackley}, {Adams}, {Adams}, {Addesso}, {Adhikari}, {Adya}, \&
  et~al.}]{2017ApJ...848L..12A}
--- 2017{\natexlab{d}}, \apjl, 848, L12

\bibitem[{{Alexander} {et~al.}(2017){Alexander}, {Berger}, {Fong}, {Williams},
  {Guidorzi}, {Margutti}, {Metzger}, {Annis}, {Blanchard}, {Brout}, {Brown},
  {Chen}, {Chornock}, {Cowperthwaite}, {Drout}, {Eftekhari}, {Frieman}, {Holz},
  {Nicholl}, {Rest}, {Sako}, {Soares-Santos}, \&
  {Villar}}]{2017ApJ...848L..21A}
{Alexander}, K.~D. {et al.}\  2017, \apjl, 848, L21

\bibitem[{{Alexander} {et~al.}(2018){Alexander}, {Margutti}, {Blanchard},
  {Fong}, {Berger}, {Hajela}, {Eftekhari}, {Chornock}, {Cowperthwaite},
  {Giannios}, {Guidorzi}, {Kathirgamaraju}, {MacFadyen}, {Metzger}, {Nicholl},
  {Sironi}, {Villar}, {Williams}, {Xie}, \& {Zrake}}]{2018arXiv180502870A}
--- 2018, \apjl, 863, L18

\bibitem[{{Aloy} {et~al.}(2005){Aloy}, {Janka}, \&
  {M{\"u}ller}}]{2005A&A...436..273A}
{Aloy}, M.~A., {Janka}, H.-T., \& {M{\"u}ller}, E. 2005, \aap, 436, 273

\bibitem[{{Amati} {et~al.}(2002){Amati}, {Frontera}, {Tavani}, {in't Zand},
  {Antonelli}, {Costa}, {Feroci}, {Guidorzi}, {Heise}, {Masetti}, {Montanari},
  {Nicastro}, {Palazzi}, {Pian}, {Piro}, \& {Soffitta}}]{2002A&A...390...81A}
{Amati}, L. {et al.}\  2002, \aap, 390, 81

\bibitem[{{Andreoni} {et~al.}(2017){Andreoni}, {Ackley}, {Cooke}, {Acharyya},
  {Allison}, {Anderson}, {Ashley}, {Baade}, {Bailes}, {Bannister}, {Beardsley},
  {Bessell}, {Bian}, {Bland}, {Boer}, {Booler}, {Brandeker}, {Brown},
  {Buckley}, {Chang}, {Coward}, {Crawford}, {Crisp}, {Crosse}, {Cucchiara},
  {Cup{\'a}k}, {de Gois}, {Deller}, {Devillepoix}, {Dobie}, {Elmer}, {Emrich},
  {Farah}, {Farrell}, {Franzen}, {Gaensler}, {Galloway}, {Gendre}, {Giblin},
  {Goobar}, {Green}, {Hancock}, {Hartig}, {Howell}, {Horsley}, {Hotan},
  {Howie}, {Hu}, {Hu}, {James}, {Johnston}, {Johnston-Hollitt}, {Kaplan},
  {Kasliwal}, {Keane}, {Kenney}, {Klotz}, {Lau}, {Laugier}, {Lenc}, {Li},
  {Liang}, {Lidman}, {Luvaul}, {Lynch}, {Ma}, {Macpherson}, {Mao},
  {McClelland}, {McCully}, {M{\"o}ller}, {Morales}, {Morris}, {Murphy},
  {Noysena}, {Onken}, {Orange}, {Os{\l}owski}, {Pallot}, {Paxman}, {Potter},
  {Pritchard}, {Raja}, {Ridden-Harper}, {Romero-Colmenero}, {Sadler}, {Sansom},
  {Scalzo}, {Schmidt}, {Scott}, {Seghouani}, {Shang}, {Shannon}, {Shao},
  {Shara}, {Sharp}, {Sokolowski}, {Sollerman}, {Staff}, {Steele}, {Sun},
  {Suntzeff}, {Tao}, {Tingay}, {Towner}, {Thierry}, {Trott}, {Tucker},
  {V{\"a}is{\"a}nen}, {Krishnan}, {Walker}, {Wang}, {Wang}, {Wayth}, {Whiting},
  {Williams}, {Williams}, {Wolf}, {Wu}, {Wu}, {Yang}, {Yuan}, {Zhang}, {Zhou},
  \& {Zovaro}}]{2017PASA...34...69A}
{Andreoni}, I. {et al.}\  2017, PASA, 34, e069

\bibitem[{{Arcavi} {et~al.}(2017){Arcavi}, {Hosseinzadeh}, {Howell}, {McCully},
  {Poznanski}, {Kasen}, {Barnes}, {Zaltzman}, {Vasylyev}, {Maoz}, \&
  {Valenti}}]{2017Natur.551...64A}
{Arcavi}, I. {et al.}\  2017, \nat, 551, 64

\bibitem[{{Band} {et~al.}(1993){Band}, {Matteson}, {Ford}, {Schaefer},
  {Palmer}, {Teegarden}, {Cline}, {Briggs}, {Paciesas}, {Pendleton}, {Fishman},
  {Kouveliotou}, {Meegan}, {Wilson}, \& {Lestrade}}]{1993ApJ...413..281B}
{Band}, D. {et al.}\  1993, \apj, 413, 281

\bibitem[{{B{\'e}gu{\'e}} {et~al.}(2017){B{\'e}gu{\'e}}, {Burgess}, \&
  {Greiner}}]{2017ApJ...851L..19B}
{B{\'e}gu{\'e}}, D., {Burgess}, J.~M., \& {Greiner}, J. 2017, \apjl, 851, L19

\bibitem[{{Beniamini} \& {Giannios}(2017)}]{2017MNRAS.468.3202B}
{Beniamini}, P. \& {Giannios}, D. 2017, \mnras, 468, 3202

\bibitem[{{Beniamini} {et~al.}(2016){Beniamini}, {Nava}, \&
  {Piran}}]{2016MNRAS.461...51B}
{Beniamini}, P., {Nava}, L., \& {Piran}, T. 2016, \mnras, 461, 51

\bibitem[{{Berger}(2014)}]{2014ARA&A..52...43B}
{Berger}, E. 2014, \araa, 52, 43

\bibitem[{{Bovard} {et~al.}(2017){Bovard}, {Martin}, {Guercilena}, {Arcones},
  {Rezzolla}, \& {Korobkin}}]{2017PhRvD..96l4005B}
{Bovard}, L., {Martin}, D., {Guercilena}, F., {Arcones}, A., {Rezzolla}, L., \&
  {Korobkin}, O. 2017, \prd, 96, 124005

\bibitem[{{Bromberg} {et~al.}(2011){Bromberg}, {Nakar}, {Piran}, \&
  {Sari}}]{2011ApJ...740..100B}
{Bromberg}, O., {Nakar}, E., {Piran}, T., \& {Sari}, R. 2011, \apj, 740, 100

\bibitem[{{Buckley} {et~al.}(2018){Buckley}, {Andreoni}, {Barway}, {Cooke},
  {Crawford}, {Gorbovskoy}, {Gromadzki}, {Lipunov}, {Mao}, {Potter},
  {Pretorius}, {Pritchard}, {Romero-Colmenero}, {Shara}, {V{\"a}is{\"a}nen}, \&
  {Williams}}]{2018MNRAS.474L..71B}
{Buckley}, D.~A.~H. {et al.}\  2018, \mnras, 474, L71

\bibitem[{{Burgess} {et~al.}(2017{\natexlab{a}}){Burgess}, {Greiner},
  {B{\'e}gu{\'e}}, \& {Berlato}}]{2017arXiv171008362B}
{Burgess}, J.~M., {Greiner}, J., {B{\'e}gu{\'e}}, D., \& {Berlato}, F.
  2017{\natexlab{a}}, arXiv:1710.08362

\bibitem[{{Burgess} {et~al.}(2017{\natexlab{b}}){Burgess}, {Greiner}, {Begue},
  {Giannios}, {Berlato}, \& {Lipunov}}]{2017arXiv171005823B}
{Burgess}, J.~M., {Greiner}, J., {Begue}, D., {Giannios}, D., {Berlato}, F., \&
  {Lipunov}, V.~M. 2017{\natexlab{b}}, arXiv:1710.05823

\bibitem[{{Coulter} {et~al.}(2017){Coulter}, {Foley}, {Kilpatrick}, {Drout},
  {Piro}, {Shappee}, {Siebert}, {Simon}, {Ulloa}, {Kasen}, {Madore},
  {Murguia-Berthier}, {Pan}, {Prochaska}, {Ramirez-Ruiz}, {Rest}, \&
  {Rojas-Bravo}}]{2017Sci...358.1556C}
{Coulter}, D.~A. {et al.}\  2017, Science, 358, 1556

\bibitem[{{Covino} {et~al.}(2017){Covino}, {Wiersema}, {Fan}, {Toma},
  {Higgins}, {Melandri}, {D'Avanzo}, {Mundell}, {Palazzi}, {Tanvir},
  {Bernardini}, {Branchesi}, {Brocato}, {Campana}, {di Serego Alighieri},
  {G{\"o}tz}, {Fynbo}, {Gao}, {Gomboc}, {Gompertz}, {Greiner}, {Hjorth}, {Jin},
  {Kaper}, {Klose}, {Kobayashi}, {Kopac}, {Kouveliotou}, {Levan}, {Mao},
  {Malesani}, {Pian}, {Rossi}, {Salvaterra}, {Starling}, {Steele},
  {Tagliaferri}, {Troja}, {van der Horst}, \& {Wijers}}]{2017NatAs...1..791C}
{Covino}, S. {et al.}\  2017, Nature Astronomy, 1, 791

\bibitem[{{Cowperthwaite} {et~al.}(2017){Cowperthwaite}, {Berger}, {Villar},
  {Metzger}, {Nicholl}, {Chornock}, {Blanchard}, {Fong}, {Margutti},
  {Soares-Santos}, {Alexander}, {Allam}, {Annis}, {Brout}, {Brown}, {Butler},
  {Chen}, {Diehl}, {Doctor}, {Drout}, {Eftekhari}, {Farr}, {Finley}, {Foley},
  {Frieman}, {Fryer}, {Garc{\'{\i}}a-Bellido}, {Gill}, {Guillochon}, {Herner},
  {Holz}, {Kasen}, {Kessler}, {Marriner}, {Matheson}, {Neilsen}, {Quataert},
  {Palmese}, {Rest}, {Sako}, {Scolnic}, {Smith}, {Tucker}, {Williams},
  {Balbinot}, {Carlin}, {Cook}, {Durret}, {Li}, {Lopes}, {Louren{\c c}o},
  {Marshall}, {Medina}, {Muir}, {Mu{\~n}oz}, {Sauseda}, {Schlegel}, {Secco},
  {Vivas}, {Wester}, {Zenteno}, {Zhang}, {Abbott}, {Banerji}, {Bechtol},
  {Benoit-L{\'e}vy}, {Bertin}, {Buckley-Geer}, {Burke}, {Capozzi}, {Carnero
  Rosell}, {Carrasco Kind}, {Castander}, {Crocce}, {Cunha}, {D'Andrea}, {da
  Costa}, {Davis}, {DePoy}, {Desai}, {Dietrich}, {Drlica-Wagner}, {Eifler},
  {Evrard}, {Fernandez}, {Flaugher}, {Fosalba}, {Gaztanaga}, {Gerdes},
  {Giannantonio}, {Goldstein}, {Gruen}, {Gruendl}, {Gutierrez}, {Honscheid},
  {Jain}, {James}, {Jeltema}, {Johnson}, {Johnson}, {Kent}, {Krause}, {Kron},
  {Kuehn}, {Nuropatkin}, {Lahav}, {Lima}, {Lin}, {Maia}, {March}, {Martini},
  {McMahon}, {Menanteau}, {Miller}, {Miquel}, {Mohr}, {Neilsen}, {Nichol},
  {Ogando}, {Plazas}, {Roe}, {Romer}, {Roodman}, {Rykoff}, {Sanchez},
  {Scarpine}, {Schindler}, {Schubnell}, {Sevilla-Noarbe}, {Smith}, {Smith},
  {Sobreira}, {Suchyta}, {Swanson}, {Tarle}, {Thomas}, {Thomas}, {Troxel},
  {Vikram}, {Walker}, {Wechsler}, {Weller}, {Yanny}, \&
  {Zuntz}}]{2017ApJ...848L..17C}
{Cowperthwaite}, P.~S. {et al.}\  2017, \apjl, 848, L17

\bibitem[{{D'Avanzo} {et~al.}(2018){D'Avanzo}, {Campana}, {Salafia},
  {Ghirlanda}, {Ghisellini}, {Melandri}, {Bernardini}, {Branchesi},
  {Chassande-Mottin}, {Covino}, {D'Elia}, {Nava}, {Salvaterra}, {Tagliaferri},
  \& {Vergani}}]{2018A&A...613L...1D}
{D'Avanzo}, P. {et al.}\  2018, \aap, 613, L1

\bibitem[{{D{\'{\i}}az} {et~al.}(2017){D{\'{\i}}az}, {Macri}, {Garcia Lambas},
  {Mendes de Oliveira}, {Nilo Castell{\'o}n}, {Ribeiro}, {S{\'a}nchez},
  {Schoenell}, {Abramo}, {Akras}, {Alcaniz}, {Artola}, {Beroiz}, {Bonoli},
  {Cabral}, {Camuccio}, {Castillo}, {Chavushyan}, {Coelho}, {Colazo},
  {Costa-Duarte}, {Cuevas Larenas}, {DePoy}, {Dom{\'{\i}}nguez Romero},
  {Dultzin}, {Fern{\'a}ndez}, {Garc{\'{\i}}a}, {Girardini}, {Gon{\c c}alves},
  {Gon{\c c}alves}, {Gurovich}, {Jim{\'e}nez-Teja}, {Kanaan}, {Lares}, {Lopes
  de Oliveira}, {L{\'o}pez-Cruz}, {Marshall}, {Melia}, {Molino}, {Padilla},
  {Pe{\~n}uela}, {Placco}, {Qui{\~n}ones}, {Ram{\'{\i}}rez Rivera}, {Renzi},
  {Riguccini}, {R{\'{\i}}os-L{\'o}pez}, {Rodriguez}, {Sampedro}, {Schneiter},
  {Sodr{\'e}}, {Starck}, {Torres-Flores}, {Tornatore}, \&
  {Zadro{\.z}ny}}]{2017ApJ...848L..29D}
{D{\'{\i}}az}, M.~C. {et al.}\  2017, \apjl, 848, L29

\bibitem[{{Dietrich} {et~al.}(2017{\natexlab{a}}){Dietrich}, {Bernuzzi},
  {Ujevic}, \& {Tichy}}]{2017PhRvD..95d4045D}
{Dietrich}, T., {Bernuzzi}, S., {Ujevic}, M., \& {Tichy}, W.
  2017{\natexlab{a}}, \prd, 95, 044045

\bibitem[{{Dietrich} {et~al.}(2017{\natexlab{b}}){Dietrich}, {Ujevic}, {Tichy},
  {Bernuzzi}, \& {Br{\"u}gmann}}]{2017PhRvD..95b4029D}
{Dietrich}, T., {Ujevic}, M., {Tichy}, W., {Bernuzzi}, S., \& {Br{\"u}gmann},
  B. 2017{\natexlab{b}}, \prd, 95, 024029

\bibitem[{{Dobie} {et~al.}(2018){Dobie}, {Kaplan}, {Murphy}, {Lenc}, {Mooley},
  {Lynch}, {Corsi}, {Frail}, {Kasliwal}, \& {Hallinan}}]{2018ApJ...858L..15D}
{Dobie}, D. {et al.}\  2018, \apjl, 858, L15

\bibitem[{{Drout} {et~al.}(2017){Drout}, {Piro}, {Shappee}, {Kilpatrick},
  {Simon}, {Contreras}, {Coulter}, {Foley}, {Siebert}, {Morrell}, {Boutsia},
  {Di Mille}, {Holoien}, {Kasen}, {Kollmeier}, {Madore}, {Monson},
  {Murguia-Berthier}, {Pan}, {Prochaska}, {Ramirez-Ruiz}, {Rest}, {Adams},
  {Alatalo}, {Ba{\~n}ados}, {Baughman}, {Beers}, {Bernstein}, {Bitsakis},
  {Campillay}, {Hansen}, {Higgs}, {Ji}, {Maravelias}, {Marshall}, {Bidin},
  {Prieto}, {Rasmussen}, {Rojas-Bravo}, {Strom}, {Ulloa},
  {Vargas-Gonz{\'a}lez}, {Wan}, \& {Whitten}}]{2017Sci...358.1570D}
{Drout}, M.~R. {et al.}\  2017, Science, 358, 1570

\bibitem[{{Eichler} \& {Levinson}(1999)}]{1999ApJ...521L.117E}
{Eichler}, D. \& {Levinson}, A. 1999, \apjl, 521, L117

\bibitem[{{Eichler} {et~al.}(1989){Eichler}, {Livio}, {Piran}, \&
  {Schramm}}]{1989Natur.340..126E}
{Eichler}, D., {Livio}, M., {Piran}, T., \& {Schramm}, D.~N. 1989, \nat, 340,
  126

\bibitem[{{Evans} {et~al.}(2017){Evans}, {Cenko}, {Kennea}, {Emery}, {Kuin},
  {Korobkin}, {Wollaeger}, {Fryer}, {Madsen}, {Harrison}, {Xu}, {Nakar},
  {Hotokezaka}, {Lien}, {Campana}, {Oates}, {Troja}, {Breeveld}, {Marshall},
  {Barthelmy}, {Beardmore}, {Burrows}, {Cusumano}, {D'A{\`i}}, {D'Avanzo},
  {D'Elia}, {de Pasquale}, {Even}, {Fontes}, {Forster}, {Garcia}, {Giommi},
  {Grefenstette}, {Gronwall}, {Hartmann}, {Heida}, {Hungerford}, {Kasliwal},
  {Krimm}, {Levan}, {Malesani}, {Melandri}, {Miyasaka}, {Nousek}, {O'Brien},
  {Osborne}, {Pagani}, {Page}, {Palmer}, {Perri}, {Pike}, {Racusin}, {Rosswog},
  {Siegel}, {Sakamoto}, {Sbarufatti}, {Tagliaferri}, {Tanvir}, \&
  {Tohuvavohu}}]{2017Sci...358.1565E}
{Evans}, P.~A. {et al.}\  2017, Science, 358, 1565

\bibitem[{{Foucart} {et~al.}(2016){Foucart}, {Haas}, {Duez}, {O'Connor}, {Ott},
  {Roberts}, {Kidder}, {Lippuner}, {Pfeiffer}, \&
  {Scheel}}]{2016PhRvD..93d4019F}
{Foucart}, F. {et al.}\  2016, \prd, 93, 044019

\bibitem[{{Fraija} {et~al.}(2017){Fraija}, {Veres}, {De Colle}, {Dichiara},
  {Barniol Duran}, {Lee}, \& {Galvan-Gamez}}]{2017arXiv171008514F}
{Fraija}, N., {Veres}, P., {De Colle}, F., {Dichiara}, S., {Barniol Duran}, R.,
  {Lee}, W.~H., \& {Galvan-Gamez}, A. 2017, arXiv:1710.08514

\bibitem[{{Freedman} \& {Waxman}(2001)}]{2001ApJ...547..922F}
{Freedman}, D.~L. \& {Waxman}, E. 2001, \apj, 547, 922

\bibitem[{{Ghirlanda} {et~al.}(2018){Ghirlanda}, {Salafia}, {Paragi},
  {Giroletti}, {Yang}, {Marcote}, {Blanchard}, {Agudo}, {An}, {Bernardini},
  {Beswick}, {Branchesi}, {Campana}, {Casadio}, {Chassande-Mottin}, {Colpi},
  {Covino}, {D'Avanzo}, {D'Elia}, {Frey}, {Gawronski}, {Ghisellini}, {Gurvits},
  {Jonker}, {van Langevelde}, {Melandri}, {Moldon}, {Nava}, {Perego},
  {Perez-Torres}, {Reynolds}, {Salvaterra}, {Tagliaferri}, {Venturi},
  {Vergani}, \& {Zhang}}]{2018arXiv180800469G}
{Ghirlanda}, G. {et al.}\  2018, ArXiv e-prints

\bibitem[{{Giannios} \& {Spruit}(2005)}]{2005A&A...430....1G}
{Giannios}, D. \& {Spruit}, H.~C. 2005, \aap, 430, 1

\bibitem[{{Goldstein} {et~al.}(2017){Goldstein}, {Veres}, {Burns}, {Briggs},
  {Hamburg}, {Kocevski}, {Wilson-Hodge}, {Preece}, {Poolakkil}, {Roberts},
  {Hui}, {Connaughton}, {Racusin}, {von Kienlin}, {Dal Canton}, {Christensen},
  {Littenberg}, {Siellez}, {Blackburn}, {Broida}, {Bissaldi}, {Cleveland},
  {Gibby}, {Giles}, {Kippen}, {McBreen}, {McEnery}, {Meegan}, {Paciesas}, \&
  {Stanbro}}]{2017arXiv171005446G}
{Goldstein}, A. {et al.}\  2017, \apjl, 848, L14

\bibitem[{{Gottlieb} {et~al.}(2018{\natexlab{a}}){Gottlieb}, {Nakar}, \&
  {Piran}}]{2018MNRAS.473..576G}
{Gottlieb}, O., {Nakar}, E., \& {Piran}, T. 2018{\natexlab{a}}, \mnras, 473,
  576

\bibitem[{{Gottlieb} {et~al.}(2018{\natexlab{b}}){Gottlieb}, {Nakar}, {Piran},
  \& {Hotokezaka}}]{2017arXiv171005896G}
{Gottlieb}, O., {Nakar}, E., {Piran}, T., \& {Hotokezaka}, K.
  2018{\natexlab{b}}, \mnras, 479, 588

\bibitem[{{Granot} {et~al.}(2017{\natexlab{a}}){Granot}, {Gill}, {Guetta}, \&
  {De Colle}}]{2017arXiv171006421G}
{Granot}, J., {Gill}, R., {Guetta}, D., \& {De Colle}, F. 2017{\natexlab{a}},
  arXiv:1710.06421

\bibitem[{{Granot} {et~al.}(2017{\natexlab{b}}){Granot}, {Guetta}, \&
  {Gill}}]{2017ApJ...850L..24G}
{Granot}, J., {Guetta}, D., \& {Gill}, R. 2017{\natexlab{b}}, \apjl, 850, L24

\bibitem[{{Granot} {et~al.}(2001){Granot}, {Miller}, {Piran}, {Suen}, \&
  {Hughes}}]{2001grba.conf..312G}
{Granot}, J., {Miller}, M., {Piran}, T., {Suen}, W.~M., \& {Hughes}, P.~A.
  2001, in Gamma-ray Bursts in the Afterglow Era, ed. E.~{Costa},
  F.~{Frontera}, \& J.~{Hjorth}, 312

\bibitem[{{Granot} {et~al.}(2002){Granot}, {Panaitescu}, {Kumar}, \&
  {Woosley}}]{2002ApJ...570L..61G}
{Granot}, J., {Panaitescu}, A., {Kumar}, P., \& {Woosley}, S.~E. 2002, \apjl,
  570, L61

\bibitem[{{Haggard} {et~al.}(2017){Haggard}, {Nynka}, {Ruan}, {Kalogera},
  {Cenko}, {Evans}, \& {Kennea}}]{2017ApJ...848L..25H}
{Haggard}, D., {Nynka}, M., {Ruan}, J.~J., {Kalogera}, V., {Cenko}, S.~B.,
  {Evans}, P., \& {Kennea}, J.~A. 2017, \apjl, 848, L25

\bibitem[{{Hallinan} {et~al.}(2017){Hallinan}, {Corsi}, {Mooley}, {Hotokezaka},
  {Nakar}, {Kasliwal}, {Kaplan}, {Frail}, {Myers}, {Murphy}, {De}, {Dobie},
  {Allison}, {Bannister}, {Bhalerao}, {Chandra}, {Clarke}, {Giacintucci}, {Ho},
  {Horesh}, {Kassim}, {Kulkarni}, {Lenc}, {Lockman}, {Lynch}, {Nichols},
  {Nissanke}, {Palliyaguru}, {Peters}, {Piran}, {Rana}, {Sadler}, \&
  {Singer}}]{2017Sci...358.1579H}
{Hallinan}, G. {et al.}\  2017, Science, 358, 1579

\bibitem[{{He} {et~al.}(2018){He}, {Tam}, \& {Shen}}]{2017arXiv171005869H}
{He}, X.-B., {Tam}, P.-H.~T., \& {Shen}, R.-F. 2018, Research in Astronomy and
  Astrophysics, 18, 043

\bibitem[{{Hjorth} {et~al.}(2017){Hjorth}, {Levan}, {Tanvir}, {Lyman},
  {Wojtak}, {Schr{\o}der}, {Mandel}, {Gall}, \& {Bruun}}]{2017ApJ...848L..31H}
{Hjorth}, J. {et al.}\  2017, \apjl, 848, L31

\bibitem[{{Hotokezaka} {et~al.}(2013){Hotokezaka}, {Kiuchi}, {Kyutoku},
  {Okawa}, {Sekiguchi}, {Shibata}, \& {Taniguchi}}]{2013PhRvD..87b4001H}
{Hotokezaka}, K., {Kiuchi}, K., {Kyutoku}, K., {Okawa}, H., {Sekiguchi}, Y.-i.,
  {Shibata}, M., \& {Taniguchi}, K. 2013, \prd, 87, 024001

\bibitem[{{Hu} {et~al.}(2017){Hu}, {Wu}, {Andreoni}, {Ashley}, {Cooke}, {Cui},
  {Du}, {Dai}, {Gu}, {Hu}, {Lu}, {Li}, {Li}, {Liang}, {Liu}, {Ma}, {Shang},
  {Sun}, {Suntzeff}, {Tao}, {Udden}, {Wang}, {Wang}, {Wen}, {Xiao}, {Su},
  {Yang}, {Yang}, {Yuan}, {Zhou}, {Zhang}, {Zhou}, \&
  {Zhu}}]{2017SciBu..62.1433H}
{Hu}, L. {et al.}\  2017, Science Bulletin, Vol.~62, No.21, p.1433-1438, 2017,
  62, 1433

\bibitem[{{Im} {et~al.}(2017){Im}, {Yoon}, {Lee}, {Lee}, {Kim}, {Lee}, {Kim},
  {Troja}, {Choi}, {Lim}, {Ko}, \& {Shim}}]{2017ApJ...849L..16I}
{Im}, M. {et al.}\  2017, \apjl, 849, L16

\bibitem[{{Ioka} \& {Nakamura}(2001)}]{2001ApJ...554L.163I}
{Ioka}, K. \& {Nakamura}, T. 2001, \apjl, 554, L163

\bibitem[{{Ioka} \& {Nakamura}(2018)}]{2017arXiv171005905I}
--- 2018, Progress of Theoretical and Experimental Physics, 2018, 043E02

\bibitem[{{Ioka} {et~al.}(2011){Ioka}, {Ohira}, {Kawanaka}, \&
  {Mizuta}}]{2011PThPh.126..555I}
{Ioka}, K., {Ohira}, Y., {Kawanaka}, N., \& {Mizuta}, A. 2011, Progress of
  Theoretical Physics, 126, 555

\bibitem[{{Jin} {et~al.}(2018){Jin}, {Li}, {Wang}, {Wang}, {He}, {Yuan},
  {Zhang}, {Zou}, {Fan}, \& {Wei}}]{2017arXiv170807008J}
{Jin}, Z.-P. {et al.}\  2018, \apj, 857, 128

\bibitem[{{Kasen} {et~al.}(2017){Kasen}, {Metzger}, {Barnes}, {Quataert}, \&
  {Ramirez-Ruiz}}]{2017Natur.551...80K}
{Kasen}, D., {Metzger}, B., {Barnes}, J., {Quataert}, E., \& {Ramirez-Ruiz}, E.
  2017, \nat, 551, 80

\bibitem[{{Kasliwal} {et~al.}(2017){Kasliwal}, {Nakar}, {Singer}, {Kaplan},
  {Cook}, {Van Sistine}, {Lau}, {Fremling}, {Gottlieb}, {Jencson}, {Adams},
  {Feindt}, {Hotokezaka}, {Ghosh}, {Perley}, {Yu}, {Piran}, {Allison},
  {Anupama}, {Balasubramanian}, {Bannister}, {Bally}, {Barnes}, {Barway},
  {Bellm}, {Bhalerao}, {Bhattacharya}, {Blagorodnova}, {Bloom}, {Brady},
  {Cannella}, {Chatterjee}, {Cenko}, {Cobb}, {Copperwheat}, {Corsi}, {De},
  {Dobie}, {Emery}, {Evans}, {Fox}, {Frail}, {Frohmaier}, {Goobar}, {Hallinan},
  {Harrison}, {Helou}, {Hinderer}, {Ho}, {Horesh}, {Ip}, {Itoh}, {Kasen},
  {Kim}, {Kuin}, {Kupfer}, {Lynch}, {Madsen}, {Mazzali}, {Miller}, {Mooley},
  {Murphy}, {Ngeow}, {Nichols}, {Nissanke}, {Nugent}, {Ofek}, {Qi}, {Quimby},
  {Rosswog}, {Rusu}, {Sadler}, {Schmidt}, {Sollerman}, {Steele}, {Williamson},
  {Xu}, {Yan}, {Yatsu}, {Zhang}, \& {Zhao}}]{2017Sci...358.1559K}
{Kasliwal}, M.~M. {et al.}\  2017, Science, 358, 1559

\bibitem[{{Kathirgamaraju} {et~al.}(2018){Kathirgamaraju}, {Barniol Duran}, \&
  {Giannios}}]{2018MNRAS.473L.121K}
{Kathirgamaraju}, A., {Barniol Duran}, R., \& {Giannios}, D. 2018, \mnras, 473,
  L121

\bibitem[{{Kilpatrick} {et~al.}(2017){Kilpatrick}, {Foley}, {Kasen},
  {Murguia-Berthier}, {Ramirez-Ruiz}, {Coulter}, {Drout}, {Piro}, {Shappee},
  {Boutsia}, {Contreras}, {Di Mille}, {Madore}, {Morrell}, {Pan}, {Prochaska},
  {Rest}, {Rojas-Bravo}, {Siebert}, {Simon}, \& {Ulloa}}]{2017Sci...358.1583K}
{Kilpatrick}, C.~D. {et al.}\  2017, Science, 358, 1583

\bibitem[{{Kim} {et~al.}(2017){Kim}, {Schulze}, {Resmi},
  {Gonz{\'a}lez-L{\'o}pez}, {Higgins}, {Ishwara-Chandra}, {Bauer}, {de
  Gregorio-Monsalvo}, {De Pasquale}, {de Ugarte Postigo}, {Kann},
  {Mart{\'{\i}}n}, {Oates}, {Starling}, {Tanvir}, {Buchner}, {Campana}, {Cano},
  {Covino}, {Fruchter}, {Fynbo}, {Hartmann}, {Hjorth}, {Jakobsson}, {Levan},
  {Malesani}, {Micha{\l}owski}, {Milvang-Jensen}, {Misra}, {O'Brien},
  {S{\'a}nchez-Ram{\'{\i}}rez}, {Th{\"o}ne}, {Watson}, \&
  {Wiersema}}]{2017ApJ...850L..21K}
{Kim}, S. {et al.}\  2017, \apjl, 850, L21

\bibitem[{{Kisaka} {et~al.}(2015){Kisaka}, {Ioka}, \&
  {Nakamura}}]{2015ApJ...809L...8K}
{Kisaka}, S., {Ioka}, K., \& {Nakamura}, T. 2015, \apjl, 809, L8

\bibitem[{{Kisaka} {et~al.}(2017){Kisaka}, {Ioka}, \&
  {Sakamoto}}]{2017ApJ...846..142K}
{Kisaka}, S., {Ioka}, K., \& {Sakamoto}, T. 2017, \apj, 846, 142

\bibitem[{{Kobayashi} {et~al.}(1997){Kobayashi}, {Piran}, \&
  {Sari}}]{1997ApJ...490...92K}
{Kobayashi}, S., {Piran}, T., \& {Sari}, R. 1997, \apj, 490, 92

\bibitem[{{Krimm} {et~al.}(2013){Krimm}, {Holland}, {Corbet}, {Pearlman},
  {Romano}, {Kennea}, {Bloom}, {Barthelmy}, {Baumgartner}, {Cummings},
  {Gehrels}, {Lien}, {Markwardt}, {Palmer}, {Sakamoto}, {Stamatikos}, \&
  {Ukwatta}}]{2013ApJS..209...14K}
{Krimm}, H.~A. {et al.}\  2013, \apjs, 209, 14

\bibitem[{{Kumar} \& {Zhang}(2015)}]{2015PhR...561....1K}
{Kumar}, P. \& {Zhang}, B. 2015, \physrep, 561, 1

\bibitem[{{Kyutoku} {et~al.}(2014){Kyutoku}, {Ioka}, \&
  {Shibata}}]{2014MNRAS.437L...6K}
{Kyutoku}, K., {Ioka}, K., \& {Shibata}, M. 2014, \mnras, 437, L6

\bibitem[{{Lamb} \& {Kobayashi}(2016)}]{2016ApJ...829..112L}
{Lamb}, G.~P. \& {Kobayashi}, S. 2016, \apj, 829, 112

\bibitem[{{Lamb} \& {Kobayashi}(2017)}]{2017MNRAS.472.4953L}
--- 2017, \mnras, 472, 4953

\bibitem[{{Lamb} \& {Kobayashi}(2018)}]{2017arXiv171005857L}
--- 2018, \mnras, 478, 733

\bibitem[{{Lazzati} {et~al.}(2017{\natexlab{a}}){Lazzati}, {Deich}, {Morsony},
  \& {Workman}}]{2017MNRAS.471.1652L}
{Lazzati}, D., {Deich}, A., {Morsony}, B.~J., \& {Workman}, J.~C.
  2017{\natexlab{a}}, \mnras, 471, 1652

\bibitem[{{Lazzati} {et~al.}(2017{\natexlab{b}}){Lazzati},
  {L{\'o}pez-C{\'a}mara}, {Cantiello}, {Morsony}, {Perna}, \&
  {Workman}}]{2017ApJ...848L...6L}
{Lazzati}, D., {L{\'o}pez-C{\'a}mara}, D., {Cantiello}, M., {Morsony}, B.~J.,
  {Perna}, R., \& {Workman}, J.~C. 2017{\natexlab{b}}, \apjl, 848, L6

\bibitem[{{Lazzati} {et~al.}(2018){Lazzati}, {Perna}, {Morsony},
  {Lopez-Camara}, {Cantiello}, {Ciolfi}, {Giacomazzo}, \&
  {Workman}}]{2017arXiv171203237L}
{Lazzati}, D., {Perna}, R., {Morsony}, B.~J., {Lopez-Camara}, D., {Cantiello},
  M., {Ciolfi}, R., {Giacomazzo}, B., \& {Workman}, J.~C. 2018, Physical Review
  Letters, 120, 241103

\bibitem[{{Lehner} {et~al.}(2016){Lehner}, {Liebling}, {Palenzuela},
  {Caballero}, {O'Connor}, {Anderson}, \& {Neilsen}}]{2016CQGra..33r4002L}
{Lehner}, L., {Liebling}, S.~L., {Palenzuela}, C., {Caballero}, O.~L.,
  {O'Connor}, E., {Anderson}, M., \& {Neilsen}, D. 2016, Classical and Quantum
  Gravity, 33, 184002

\bibitem[{{Lipunov} {et~al.}(2017){Lipunov}, {Gorbovskoy}, {Kornilov},
  {.~Tyurina}, {Balanutsa}, {Kuznetsov}, {Vlasenko}, {Kuvshinov}, {Gorbunov},
  {Buckley}, {Krylov}, {Podesta}, {Lopez}, {Podesta}, {Levato}, {Saffe},
  {Mallamachi}, {Potter}, {Budnev}, {Gress}, {Ishmuhametova}, {Vladimirov},
  {Zimnukhov}, {Yurkov}, {Sergienko}, {Gabovich}, {Rebolo}, {Serra-Ricart},
  {Israelyan}, {Chazov}, {Wang}, {Tlatov}, \&
  {Panchenko}}]{2017ApJ...850L...1L}
{Lipunov}, V.~M. {et al.}\  2017, \apjl, 850, L1

\bibitem[{{Lithwick} \& {Sari}(2001)}]{2001ApJ...555..540L}
{Lithwick}, Y. \& {Sari}, R. 2001, \apj, 555, 540

\bibitem[{{Lyman} {et~al.}(2018){Lyman}, {Lamb}, {Levan}, {Mandel}, {Tanvir},
  {Kobayashi}, {Gompertz}, {Hjorth}, {Fruchter}, {Kangas}, {Steeghs}, {Steele},
  {Cano}, {Copperwheat}, {Evans}, {Fynbo}, {Gall}, {Im}, {Izzo}, {Jakobsson},
  {Milvang-Jensen}, {O'Brien}, {Osborne}, {Palazzi}, {Perley}, {Pian},
  {Rosswog}, {Rowlinson}, {Schulze}, {Stanway}, {Sutton}, {Th{\"o}ne}, {de
  Ugarte Postigo}, {Watson}, {Wiersema}, \& {Wijers}}]{2018arXiv180102669L}
{Lyman}, J.~D. {et al.}\  2018, Nature Astronomy, 2, 751

\bibitem[{{Margutti} {et~al.}(2018){Margutti}, {Alexander}, {Xie}, {Sironi},
  {Metzger}, {Kathirgamaraju}, {Fong}, {Blanchard}, {Berger}, {MacFadyen},
  {Giannios}, {Guidorzi}, {Hajela}, {Chornock}, {Cowperthwaite}, {Eftekhari},
  {Nicholl}, {Villar}, {Williams}, \& {Zrake}}]{2018ApJ...856L..18M}
{Margutti}, R. {et al.}\  2018, \apjl, 856, L18

\bibitem[{{Margutti} {et~al.}(2017){Margutti}, {Berger}, {Fong}, {Guidorzi},
  {Alexander}, {Metzger}, {Blanchard}, {Cowperthwaite}, {Chornock},
  {Eftekhari}, {Nicholl}, {Villar}, {Williams}, {Annis}, {Brown}, {Chen},
  {Doctor}, {Frieman}, {Holz}, {Sako}, \&
  {Soares-Santos}}]{2017ApJ...848L..20M}
--- 2017, \apjl, 848, L20

\bibitem[{{Matsumoto} {et~al.}(2018){Matsumoto}, {Ioka}, {Kisaka}, \&
  {Nakar}}]{2018arXiv180207732M}
{Matsumoto}, T., {Ioka}, K., {Kisaka}, S., \& {Nakar}, E. 2018, \apj, 861, 55

\bibitem[{{McCully} {et~al.}(2017){McCully}, {Hiramatsu}, {Howell},
  {Hosseinzadeh}, {Arcavi}, {Kasen}, {Barnes}, {Shara}, {Williams},
  {V{\"a}is{\"a}nen}, {Potter}, {Romero-Colmenero}, {Crawford}, {Buckley},
  {Cooke}, {Andreoni}, {Pritchard}, {Mao}, {Gromadzki}, \&
  {Burke}}]{2017ApJ...848L..32M}
{McCully}, C. {et al.}\  2017, \apjl, 848, L32

\bibitem[{{M{\'e}sz{\'a}ros}(2006)}]{2006RPPh...69.2259M}
{M{\'e}sz{\'a}ros}, P. 2006, Reports on Progress in Physics, 69, 2259

\bibitem[{{Meszaros} \& {Rees}(1992)}]{1992ApJ...397..570M}
{Meszaros}, P. \& {Rees}, M.~J. 1992, \apj, 397, 570

\bibitem[{{M{\'e}sz{\'a}ros} \& {Rees}(2000)}]{2000ApJ...530..292M}
{M{\'e}sz{\'a}ros}, P. \& {Rees}, M.~J. 2000, \apj, 530, 292

\bibitem[{{Mizuta} \& {Ioka}(2013)}]{2013ApJ...777..162M}
{Mizuta}, A. \& {Ioka}, K. 2013, \apj, 777, 162

\bibitem[{{Mooley} {et~al.}(2018{\natexlab{a}}){Mooley}, {Deller}, {Gottlieb},
  {Nakar}, {Hallinan}, {Bourke}, {Frail}, {Horesh}, {Corsi}, \&
  {Hotokezaka}}]{2018arXiv180609693M}
{Mooley}, K.~P. {et al.}\  2018{\natexlab{a}}, ArXiv e-prints

\bibitem[{{Mooley} {et~al.}(2018{\natexlab{b}}){Mooley}, {Nakar}, {Hotokezaka},
  {Hallinan}, {Corsi}, {Frail}, {Horesh}, {Murphy}, {Lenc}, {Kaplan}, {de},
  {Dobie}, {Chandra}, {Deller}, {Gottlieb}, {Kasliwal}, {Kulkarni}, {Myers},
  {Nissanke}, {Piran}, {Lynch}, {Bhalerao}, {Bourke}, {Bannister}, \&
  {Singer}}]{2018Natur.554..207M}
--- 2018{\natexlab{b}}, \nat, 554, 207

\bibitem[{{Murase} \& {Ioka}(2008)}]{2008ApJ...676.1123M}
{Murase}, K. \& {Ioka}, K. 2008, \apj, 676, 1123

\bibitem[{{Murguia-Berthier} {et~al.}(2014){Murguia-Berthier}, {Montes},
  {Ramirez-Ruiz}, {De Colle}, \& {Lee}}]{2014ApJ...788L...8M}
{Murguia-Berthier}, A., {Montes}, G., {Ramirez-Ruiz}, E., {De Colle}, F., \&
  {Lee}, W.~H. 2014, \apjl, 788, L8

\bibitem[{{Murguia-Berthier} {et~al.}(2017{\natexlab{a}}){Murguia-Berthier},
  {Ramirez-Ruiz}, {Kilpatrick}, {Foley}, {Kasen}, {Lee}, {Piro}, {Coulter},
  {Drout}, {Madore}, {Shappee}, {Pan}, {Prochaska}, {Rest}, {Rojas-Bravo},
  {Siebert}, \& {Simon}}]{2017ApJ...848L..34M}
{Murguia-Berthier}, A. {et al.}\  2017{\natexlab{a}}, \apjl, 848, L34

\bibitem[{{Murguia-Berthier} {et~al.}(2017{\natexlab{b}}){Murguia-Berthier},
  {Ramirez-Ruiz}, {Montes}, {De Colle}, {Rezzolla}, {Rosswog}, {Takami},
  {Perego}, \& {Lee}}]{2017ApJ...835L..34M}
--- 2017{\natexlab{b}}, \apjl, 835, L34

\bibitem[{{Nagakura} {et~al.}(2014){Nagakura}, {Hotokezaka}, {Sekiguchi},
  {Shibata}, \& {Ioka}}]{2014ApJ...784L..28N}
{Nagakura}, H., {Hotokezaka}, K., {Sekiguchi}, Y., {Shibata}, M., \& {Ioka}, K.
  2014, \apjl, 784, L28

\bibitem[{{Nakamura}(1998)}]{1998PThPh.100..921N}
{Nakamura}, T. 1998, Progress of Theoretical Physics, 100, 921

\bibitem[{{Nakar} \& {Piran}(2017)}]{2017ApJ...834...28N}
{Nakar}, E. \& {Piran}, T. 2017, \apj, 834, 28

\bibitem[{{Narayan} {et~al.}(1992){Narayan}, {Paczynski}, \&
  {Piran}}]{1992ApJ...395L..83N}
{Narayan}, R., {Paczynski}, B., \& {Piran}, T. 1992, \apjl, 395, L83

\bibitem[{{Narayana Bhat} {et~al.}(2016){Narayana Bhat}, {Meegan}, {von
  Kienlin}, {Paciesas}, {Briggs}, {Burgess}, {Burns}, {Chaplin}, {Cleveland},
  {Collazzi}, {Connaughton}, {Diekmann}, {Fitzpatrick}, {Gibby}, {Giles},
  {Goldstein}, {Greiner}, {Jenke}, {Kippen}, {Kouveliotou}, {Mailyan},
  {McBreen}, {Pelassa}, {Preece}, {Roberts}, {Sparke}, {Stanbro}, {Veres},
  {Wilson-Hodge}, {Xiong}, {Younes}, {Yu}, \& {Zhang}}]{2016ApJS..223...28N}
{Narayana Bhat}, P. {et al.}\  2016, \apjs, 223, 28

\bibitem[{{Nicholl} {et~al.}(2017){Nicholl}, {Berger}, {Kasen}, {Metzger},
  {Elias}, {Brice{\~n}o}, {Alexander}, {Blanchard}, {Chornock},
  {Cowperthwaite}, {Eftekhari}, {Fong}, {Margutti}, {Villar}, {Williams},
  {Brown}, {Annis}, {Bahramian}, {Brout}, {Brown}, {Chen}, {Clemens},
  {Dennihy}, {Dunlap}, {Holz}, {Marchesini}, {Massaro}, {Moskowitz},
  {Pelisoli}, {Rest}, {Ricci}, {Sako}, {Soares-Santos}, \&
  {Strader}}]{2017ApJ...848L..18N}
{Nicholl}, M. {et al.}\  2017, \apjl, 848, L18

\bibitem[{{Nynka} {et~al.}(2018){Nynka}, {Ruan}, {Haggard}, \&
  {Evans}}]{2018arXiv180504093N}
{Nynka}, M., {Ruan}, J.~J., {Haggard}, D., \& {Evans}, P.~A. 2018, \apjl, 862,
  L19

\bibitem[{{Paczynski}(1986)}]{1986ApJ...308L..43P}
{Paczynski}, B. 1986, \apjl, 308, L43

\bibitem[{{Paczynski} \& {Xu}(1994)}]{1994ApJ...427..708P}
{Paczynski}, B. \& {Xu}, G. 1994, \apj, 427, 708

\bibitem[{{Pian} {et~al.}(2017){Pian}, {D'Avanzo}, {Benetti}, {Branchesi},
  {Brocato}, {Campana}, {Cappellaro}, {Covino}, {D'Elia}, {Fynbo}, {Getman},
  {Ghirlanda}, {Ghisellini}, {Grado}, {Greco}, {Hjorth}, {Kouveliotou},
  {Levan}, {Limatola}, {Malesani}, {Mazzali}, {Melandri}, {M{\o}ller},
  {Nicastro}, {Palazzi}, {Piranomonte}, {Rossi}, {Salafia}, {Selsing},
  {Stratta}, {Tanaka}, {Tanvir}, {Tomasella}, {Watson}, {Yang}, {Amati},
  {Antonelli}, {Ascenzi}, {Bernardini}, {Bo{\"e}r}, {Bufano}, {Bulgarelli},
  {Capaccioli}, {Casella}, {Castro-Tirado}, {Chassande-Mottin}, {Ciolfi},
  {Copperwheat}, {Dadina}, {De Cesare}, {di Paola}, {Fan}, {Gendre},
  {Giuffrida}, {Giunta}, {Hunt}, {Israel}, {Jin}, {Kasliwal}, {Klose}, {Lisi},
  {Longo}, {Maiorano}, {Mapelli}, {Masetti}, {Nava}, {Patricelli}, {Perley},
  {Pescalli}, {Piran}, {Possenti}, {Pulone}, {Razzano}, {Salvaterra},
  {Schipani}, {Spera}, {Stamerra}, {Stella}, {Tagliaferri}, {Testa}, {Troja},
  {Turatto}, {Vergani}, \& {Vergani}}]{2017Natur.551...67P}
{Pian}, E. {et al.}\  2017, \nat, 551, 67

\bibitem[{{Piro} \& {Kollmeier}(2018)}]{2018ApJ...855..103P}
{Piro}, A.~L. \& {Kollmeier}, J.~A. 2018, \apj, 855, 103

\bibitem[{{Pooley} {et~al.}(2018){Pooley}, {Kumar}, {Wheeler}, \&
  {Grossan}}]{2018ApJ...859L..23P}
{Pooley}, D., {Kumar}, P., {Wheeler}, J.~C., \& {Grossan}, B. 2018, \apjl, 859,
  L23

\bibitem[{{Pozanenko} {et~al.}(2018){Pozanenko}, {Barkov}, {Minaev}, {Volnova},
  {Mazaeva}, {Moskvitin}, {Krugov}, {Samodurov}, {Loznikov}, \&
  {Lyutikov}}]{2018ApJ...852L..30P}
{Pozanenko}, A.~S. {et al.}\  2018, \apjl, 852, L30

\bibitem[{{Pozdnyakov} {et~al.}(1983){Pozdnyakov}, {Sobol}, \&
  {Syunyaev}}]{1983ASPRv...2..189P}
{Pozdnyakov}, L.~A., {Sobol}, I.~M., \& {Syunyaev}, R.~A. 1983, Astrophysics
  and Space Physics Reviews, 2, 189

\bibitem[{{Radice} {et~al.}(2016){Radice}, {Galeazzi}, {Lippuner}, {Roberts},
  {Ott}, \& {Rezzolla}}]{2016MNRAS.460.3255R}
{Radice}, D., {Galeazzi}, F., {Lippuner}, J., {Roberts}, L.~F., {Ott}, C.~D.,
  \& {Rezzolla}, L. 2016, \mnras, 460, 3255

\bibitem[{{Rees} \& {Meszaros}(1994)}]{1994ApJ...430L..93R}
{Rees}, M.~J. \& {Meszaros}, P. 1994, \apjl, 430, L93

\bibitem[{{Resmi} {et~al.}(2018){Resmi}, {Schulze}, {Ishwara Chandra}, {Misra},
  {Buchner}, {De Pasquale}, {Sanchez Ramirez}, {Klose}, {Kim}, {Tanvir}, \&
  {O'Brien}}]{2018arXiv180302768R}
{Resmi}, L. {et al.}\  2018, ArXiv e-prints

\bibitem[{{Ruan} {et~al.}(2018){Ruan}, {Nynka}, {Haggard}, {Kalogera}, \&
  {Evans}}]{2018ApJ...853L...4R}
{Ruan}, J.~J., {Nynka}, M., {Haggard}, D., {Kalogera}, V., \& {Evans}, P. 2018,
  \apjl, 853, L4

\bibitem[{{Rybicki} \& {Lightman}(1979)}]{1979rpa..book.....R}
{Rybicki}, G.~B. \& {Lightman}, A.~P. 1979, {Radiative processes in
  astrophysics}

\bibitem[{{Sari}(1997)}]{1997ApJ...489L..37S}
{Sari}, R. 1997, \apjl, 489, L37

\bibitem[{{Savchenko} {et~al.}(2017){Savchenko}, {Ferrigno}, {Kuulkers},
  {Bazzano}, {Bozzo}, {Brandt}, {Chenevez}, {Courvoisier}, {Diehl}, {Domingo},
  {Hanlon}, {Jourdain}, {von Kienlin}, {Laurent}, {Lebrun}, {Lutovinov},
  {Martin-Carrillo}, {Mereghetti}, {Natalucci}, {Rodi}, {Roques}, {Sunyaev}, \&
  {Ubertini}}]{2017arXiv171005449S}
{Savchenko}, V. {et al.}\  2017, \apjl, 848, L15

\bibitem[{{Sekiguchi} {et~al.}(2015){Sekiguchi}, {Kiuchi}, {Kyutoku}, \&
  {Shibata}}]{2015PhRvD..91f4059S}
{Sekiguchi}, Y., {Kiuchi}, K., {Kyutoku}, K., \& {Shibata}, M. 2015, \prd, 91,
  064059

\bibitem[{{Shappee} {et~al.}(2017){Shappee}, {Simon}, {Drout}, {Piro},
  {Morrell}, {Prieto}, {Kasen}, {Holoien}, {Kollmeier}, {Kelson}, {Coulter},
  {Foley}, {Kilpatrick}, {Siebert}, {Madore}, {Murguia-Berthier}, {Pan},
  {Prochaska}, {Ramirez-Ruiz}, {Rest}, {Adams}, {Alatalo}, {Ba{\~n}ados},
  {Baughman}, {Bernstein}, {Bitsakis}, {Boutsia}, {Bravo}, {Di Mille}, {Higgs},
  {Ji}, {Maravelias}, {Marshall}, {Placco}, {Prieto}, \&
  {Wan}}]{2017Sci...358.1574S}
{Shappee}, B.~J. {et al.}\  2017, Science, 358, 1574

\bibitem[{{Shibata} {et~al.}(2017){Shibata}, {Fujibayashi}, {Hotokezaka},
  {Kiuchi}, {Kyutoku}, {Sekiguchi}, \& {Tanaka}}]{2017PhRvD..96l3012S}
{Shibata}, M., {Fujibayashi}, S., {Hotokezaka}, K., {Kiuchi}, K., {Kyutoku},
  K., {Sekiguchi}, Y., \& {Tanaka}, M. 2017, \prd, 96, 123012

\bibitem[{{Siebert} {et~al.}(2017){Siebert}, {Foley}, {Drout}, {Kilpatrick},
  {Shappee}, {Coulter}, {Kasen}, {Madore}, {Murguia-Berthier}, {Pan}, {Piro},
  {Prochaska}, {Ramirez-Ruiz}, {Rest}, {Contreras}, {Morrell}, {Rojas-Bravo},
  \& {Simon}}]{2017ApJ...848L..26S}
{Siebert}, M.~R. {et al.}\  2017, \apjl, 848, L26

\bibitem[{{Smartt} {et~al.}(2017){Smartt}, {Chen}, {Jerkstrand}, {Coughlin},
  {Kankare}, {Sim}, {Fraser}, {Inserra}, {Maguire}, {Chambers}, {Huber},
  {Kr{\"u}hler}, {Leloudas}, {Magee}, {Shingles}, {Smith}, {Young}, {Tonry},
  {Kotak}, {Gal-Yam}, {Lyman}, {Homan}, {Agliozzo}, {Anderson}, {Angus},
  {Ashall}, {Barbarino}, {Bauer}, {Berton}, {Botticella}, {Bulla}, {Bulger},
  {Cannizzaro}, {Cano}, {Cartier}, {Cikota}, {Clark}, {De Cia}, {Della Valle},
  {Denneau}, {Dennefeld}, {Dessart}, {Dimitriadis}, {Elias-Rosa}, {Firth},
  {Flewelling}, {Fl{\"o}rs}, {Franckowiak}, {Frohmaier}, {Galbany},
  {Gonz{\'a}lez-Gait{\'a}n}, {Greiner}, {Gromadzki}, {Guelbenzu},
  {Guti{\'e}rrez}, {Hamanowicz}, {Hanlon}, {Harmanen}, {Heintz}, {Heinze},
  {Hernandez}, {Hodgkin}, {Hook}, {Izzo}, {James}, {Jonker}, {Kerzendorf},
  {Klose}, {Kostrzewa-Rutkowska}, {Kowalski}, {Kromer}, {Kuncarayakti},
  {Lawrence}, {Lowe}, {Magnier}, {Manulis}, {Martin-Carrillo}, {Mattila},
  {McBrien}, {M{\"u}ller}, {Nordin}, {O'Neill}, {Onori}, {Palmerio},
  {Pastorello}, {Patat}, {Pignata}, {Podsiadlowski}, {Pumo}, {Prentice}, {Rau},
  {Razza}, {Rest}, {Reynolds}, {Roy}, {Ruiter}, {Rybicki}, {Salmon}, {Schady},
  {Schultz}, {Schweyer}, {Seitenzahl}, {Smith}, {Sollerman}, {Stalder},
  {Stubbs}, {Sullivan}, {Szegedi}, {Taddia}, {Taubenberger}, {Terreran}, {van
  Soelen}, {Vos}, {Wainscoat}, {Walton}, {Waters}, {Weiland}, {Willman},
  {Wiseman}, {Wright}, {Wyrzykowski}, \& {Yaron}}]{2017Natur.551...75S}
{Smartt}, S.~J. {et al.}\  2017, \nat, 551, 75

\bibitem[{{Soares-Santos} {et~al.}(2017){Soares-Santos}, {Holz}, {Annis},
  {Chornock}, {Herner}, {Berger}, {Brout}, {Chen}, {Kessler}, {Sako}, {Allam},
  {Tucker}, {Butler}, {Palmese}, {Doctor}, {Diehl}, {Frieman}, {Yanny}, {Lin},
  {Scolnic}, {Cowperthwaite}, {Neilsen}, {Marriner}, {Kuropatkin}, {Hartley},
  {Paz-Chinch{\'o}n}, {Alexander}, {Balbinot}, {Blanchard}, {Brown}, {Carlin},
  {Conselice}, {Cook}, {Drlica-Wagner}, {Drout}, {Durret}, {Eftekhari}, {Farr},
  {Finley}, {Foley}, {Fong}, {Fryer}, {Garc{\'{\i}}a-Bellido}, {Gill},
  {Gruendl}, {Hanna}, {Kasen}, {Li}, {Lopes}, {Louren{\c c}o}, {Margutti},
  {Marshall}, {Matheson}, {Medina}, {Metzger}, {Mu{\~n}oz}, {Muir}, {Nicholl},
  {Quataert}, {Rest}, {Sauseda}, {Schlegel}, {Secco}, {Sobreira}, {Stebbins},
  {Villar}, {Vivas}, {Walker}, {Wester}, {Williams}, {Zenteno}, {Zhang},
  {Abbott}, {Abdalla}, {Banerji}, {Bechtol}, {Benoit-L{\'e}vy}, {Bertin},
  {Brooks}, {Buckley-Geer}, {Burke}, {Carnero Rosell}, {Carrasco Kind},
  {Carretero}, {Castander}, {Crocce}, {Cunha}, {D'Andrea}, {da Costa}, {Davis},
  {Desai}, {Dietrich}, {Doel}, {Eifler}, {Fernandez}, {Flaugher}, {Fosalba},
  {Gaztanaga}, {Gerdes}, {Giannantonio}, {Goldstein}, {Gruen}, {Gschwend},
  {Gutierrez}, {Honscheid}, {Jain}, {James}, {Jeltema}, {Johnson}, {Johnson},
  {Kent}, {Krause}, {Kron}, {Kuehn}, {Kuhlmann}, {Lahav}, {Lima}, {Maia},
  {March}, {McMahon}, {Menanteau}, {Miquel}, {Mohr}, {Nichol}, {Nord},
  {Ogando}, {Petravick}, {Plazas}, {Romer}, {Roodman}, {Rykoff}, {Sanchez},
  {Scarpine}, {Schubnell}, {Sevilla-Noarbe}, {Smith}, {Smith}, {Suchyta},
  {Swanson}, {Tarle}, {Thomas}, {Thomas}, {Troxel}, {Vikram}, {Wechsler},
  {Weller}, {Dark Energy Survey}, \& {Dark Energy Camera GW-EM
  Collaboration}}]{2017ApJ...848L..16S}
{Soares-Santos}, M. {et al.}\  2017, \apjl, 848, L16

\bibitem[{{Tanaka} {et~al.}(2017){Tanaka}, {Utsumi}, {Mazzali}, {Tominaga},
  {Yoshida}, {Sekiguchi}, {Morokuma}, {Motohara}, {Ohta}, {Kawabata}, {Abe},
  {Aoki}, {Asakura}, {Baar}, {Barway}, {Bond}, {Doi}, {Fujiyoshi}, {Furusawa},
  {Honda}, {Itoh}, {Kawabata}, {Kawai}, {Kim}, {Lee}, {Miyazaki}, {Morihana},
  {Nagashima}, {Nagayama}, {Nakaoka}, {Nakata}, {Ohsawa}, {Ohshima}, {Okita},
  {Saito}, {Sumi}, {Tajitsu}, {Takahashi}, {Takayama}, {Tamura}, {Tanaka},
  {Terai}, {Tristram}, {Yasuda}, \& {Zenko}}]{2017PASJ...69..102T}
{Tanaka}, M. {et al.}\  2017, \pasj, 69, 102

\bibitem[{{Tominaga} {et~al.}(2018){Tominaga}, {Tanaka}, {Morokuma}, {Utsumi},
  {Yamaguchi}, {Yasuda}, {Tanaka}, {Yoshida}, {Fujiyoshi}, {Furusawa},
  {Kawabata}, {Lee}, {Motohara}, {Ohsawa}, {Ohta}, {Terai}, {Abe}, {Aoki},
  {Asakura}, {Barway}, {Bond}, {Fujisawa}, {Honda}, {Ioka}, {Itoh}, {Kawai},
  {Kim}, {Koshimoto}, {Matsubayashi}, {Miyazaki}, {Saito}, {Sekiguchi}, {Sumi},
  \& {Tristram}}]{2018PASJ..tmp...22T}
{Tominaga}, N. {et al.}\  2018, \pasj

\bibitem[{{Troja} {et~al.}(2018){Troja}, {Piro}, {Ryan}, {van Eerten}, {Ricci},
  {Wieringa}, {Lotti}, {Sakamoto}, \& {Cenko}}]{2018MNRAS.478L..18T}
{Troja}, E. {et al.}\  2018, \mnras, 478, L18

\bibitem[{{Troja} {et~al.}(2017){Troja}, {Piro}, {van Eerten}, {Wollaeger},
  {Im}, {Fox}, {Butler}, {Cenko}, {Sakamoto}, {Fryer}, {Ricci}, {Lien}, {Ryan},
  {Korobkin}, {Lee}, {Burgess}, {Lee}, {Watson}, {Choi}, {Covino}, {D'Avanzo},
  {Fontes}, {Gonz{\'a}lez}, {Khandrika}, {Kim}, {Kim}, {Lee}, {Lee}, {Kutyrev},
  {Lim}, {S{\'a}nchez-Ram{\'{\i}}rez}, {Veilleux}, {Wieringa}, \&
  {Yoon}}]{2017Natur.551...71T}
--- 2017, \nat, 551, 71

\bibitem[{{Tsutsui} {et~al.}(2013){Tsutsui}, {Yonetoku}, {Nakamura},
  {Takahashi}, \& {Morihara}}]{2013MNRAS.431.1398T}
{Tsutsui}, R., {Yonetoku}, D., {Nakamura}, T., {Takahashi}, K., \& {Morihara},
  Y. 2013, \mnras, 431, 1398

\bibitem[{{Utsumi} {et~al.}(2017){Utsumi}, {Tanaka}, {Tominaga}, {Yoshida},
  {Barway}, {Nagayama}, {Zenko}, {Aoki}, {Fujiyoshi}, {Furusawa}, {Kawabata},
  {Koshida}, {Lee}, {Morokuma}, {Motohara}, {Nakata}, {Ohsawa}, {Ohta},
  {Okita}, {Tajitsu}, {Tanaka}, {Terai}, {Yasuda}, {Abe}, {Asakura}, {Bond},
  {Miyazaki}, {Sumi}, {Tristram}, {Honda}, {Itoh}, {Itoh}, {Kawabata},
  {Morihana}, {Nagashima}, {Nakaoka}, {Ohshima}, {Takahashi}, {Takayama},
  {Aoki}, {Baar}, {Doi}, {Finet}, {Kanda}, {Kawai}, {Kim}, {Kuroda}, {Liu},
  {Matsubayashi}, {Murata}, {Nagai}, {Saito}, {Saito}, {Sako}, {Sekiguchi},
  {Tamura}, {Tanaka}, {Uemura}, \& {Yamaguchi}}]{2017PASJ...69..101U}
{Utsumi}, Y. {et al.}\  2017, \pasj, 69, 101

\bibitem[{{Valenti} {et~al.}(2017){Valenti}, {David}, {Sand}, {Yang},
  {Cappellaro}, {Tartaglia}, {Corsi}, {Jha}, {Reichart}, {Haislip}, \&
  {Kouprianov}}]{2017ApJ...848L..24V}
{Valenti}, S. {et al.}\  2017, \apjl, 848, L24

\bibitem[{{von Kienlin} {et~al.}(2017){von Kienlin}, {Meegan}, \&
  {Goldstein}}]{2017GCN.21520....1V}
{von Kienlin}, A., {Meegan}, C., \& {Goldstein}, A. 2017, GRB Coordinates
  Network, Circular Service, No.~21520, \#1 (2017), 1520

\bibitem[{{Xiao} {et~al.}(2017){Xiao}, {Liu}, {Dai}, \&
  {Wu}}]{2017ApJ...850L..41X}
{Xiao}, D., {Liu}, L.-D., {Dai}, Z.-G., \& {Wu}, X.-F. 2017, \apjl, 850, L41

\bibitem[{{Xie} {et~al.}(2018){Xie}, {Zrake}, \&
  {MacFadyen}}]{2018arXiv180409345X}
{Xie}, X., {Zrake}, J., \& {MacFadyen}, A. 2018, \apj, 863, 58

\bibitem[{{Yonetoku} {et~al.}(2004){Yonetoku}, {Murakami}, {Nakamura},
  {Yamazaki}, {Inoue}, \& {Ioka}}]{2004ApJ...609..935Y}
{Yonetoku}, D., {Murakami}, T., {Nakamura}, T., {Yamazaki}, R., {Inoue}, A.~K.,
  \& {Ioka}, K. 2004, \apj, 609, 935

\bibitem[{{Yue} {et~al.}(2018){Yue}, {Hu}, {Zhang}, {Liang}, {Jin}, {Zou},
  {Fan}, \& {Wei}}]{2018ApJ...853L..10Y}
{Yue}, C., {Hu}, Q., {Zhang}, F.-W., {Liang}, Y.-F., {Jin}, Z.-P., {Zou},
  Y.-C., {Fan}, Y.-Z., \& {Wei}, D.-M. 2018, \apjl, 853, L10

\bibitem[{{Zhang} \& {M{\'e}sz{\'a}ros}(2001)}]{2001ApJ...559..110Z}
{Zhang}, B. \& {M{\'e}sz{\'a}ros}, P. 2001, \apj, 559, 110

\bibitem[{{Zhang} {et~al.}(2018){Zhang}, {Zhang}, {Sun}, {Lei}, {Gao}, {Li},
  {Shao}, {Zhao}, {Hu}, {L{\"u}}, {Wu}, {Fan}, {Wang}, {Castro-Tirado},
  {Zhang}, {Yu}, {Cao}, \& {Liang}}]{2018NatCo...9..447Z}
{Zhang}, B.-B. {et al.}\  2018, Nature Communications, 9, 447

\bibitem[{{Zhang} {et~al.}(2003){Zhang}, {Woosley}, \&
  {MacFadyen}}]{2003ApJ...586..356Z}
{Zhang}, W., {Woosley}, S.~E., \& {MacFadyen}, A.~I. 2003, \apj, 586, 356

\bibitem[{{Zou} {et~al.}(2018){Zou}, {Wang}, {Moharana}, {Liao}, {Chen}, {Wu},
  {Lei}, \& {Wang}}]{2018ApJ...852L...1Z}
{Zou}, Y.-C., {Wang}, F.-F., {Moharana}, R., {Liao}, B., {Chen}, W., {Wu}, Q.,
  {Lei}, W.-H., \& {Wang}, F.-Y. 2018, \apjl, 852, L1

\end{thebibliography}

\end{document}